\newif\ifconf
	\conftrue
	\ifconf
	 \documentclass[letterpaper,conference,10pt]{ieeeconf}
	 
	 \newenvironment{IEEEkeywords}{\begin{keywords}}{\end{keywords}}
	 \newcommand\IEEEQED\QED
	 
	\else
	 \documentclass[letterpaper,10pt,journal,twoside]{IEEEtran}
	\fi

	\let\labelindent\relax
	
	\usepackage{algorithm}
	\usepackage{algpseudocode}

	\usepackage{multirow}
	
	\usepackage{amsthm}
	\usepackage{amssymb}
	\usepackage{amsmath}
	\usepackage{mathrsfs}
	\usepackage{mathtools}
	
	\usepackage{bm}
	\usepackage{cite}
	
	\usepackage{graphicx}
	\usepackage{subfigure}
	\usepackage[usenames]{color}
	\usepackage{xcolor}
	
	\usepackage{balance}
	
	\usepackage{enumitem}
	\usepackage{makecell}
	\usepackage{scalerel}
	\usepackage[font={small}]{caption}
	
	\usepackage{accents}
	
	\DeclareMathAlphabet{\mathpzc}{OT1}{pzc}{m}{it}
	\DeclarePairedDelimiterX{\norm}[1]{\lVert}{\rVert}{#1}
	
	\newtheorem{lemma}{\bf{Lemma}}
	
	\newtheorem{thm}{Theorem}
	\newtheorem{remak}{Remark}

	\algnewcommand\Input{\item[\hspace{6pt}\textbf{Input:}]}
	\algnewcommand\Output{\item[\hspace{6pt}\textbf{Output:}]}
	\algnewcommand\OutputVal{\textbf{output} }
	\newcolumntype{L}[1]{>{\raggedright\arraybackslash}p{#1}}
	\newcolumntype{C}[1]{>{\centering\arraybackslash}p{#1}}
	\newcolumntype{R}[1]{>{\raggedleft\arraybackslash}p{#1}}
	
	\newcommand{\N}{\mathbb{N}}
	\newcommand{\R}{\mathbb{R}}
	\newcommand{\C}{\mathbb{C}}

	\newcommand{\cE}{\mathcal E}
	\DeclareMathOperator{\tr}{tr}
	\newcommand{\cP}{\mathcal P}

	\newcommand{\bL}{\mathbf L}
	\newcommand{\bM}{\mathbf M}

	\newcommand{\cL}{\mathcal L}

	\DeclareMathOperator{\diag}{diag}
	\DeclareMathOperator{\re}{Re}
	\DeclareMathOperator{\im}{Im}
	
	\DeclareMathOperator{\trace}{tr}
	
	\title{Distributed Koopman Operator Learning for Perception and Safe Navigation}   
	\author{Ali Azarbahram, Shenyu Liu and Gian Paolo Incremona, \IEEEmembership{Senior Member, IEEE}
	\thanks{This research was funded by the Italian Ministry of Enterprises and Made in Italy for the project 4DDS (4D Drone Swarms) under grant no. F/310097/01-04/X56, and the National Natural Science Foundation of China under grant no. 62203053.}
	\thanks{A. Azarbahram is with The Department of Electrical Engineering, Chalmers University of
	Technology, Gothenburg, 412 96, Sweden. (e-mail: ali.azarbahram@chalmers.se). S. Liu is with the School of Automation, Beijing Institute of Technology, China. (e-mail: shenyuliu@bit.edu.cn). G. P. Incremona is with the Dipartimento di Elettronica, Informazione
	e Bioingegneria, Politecnico di Milano, 20133 Milan, Italy. (e-mails: gianpaolo.incremona@polimi.it)}
	}
	
	\IEEEoverridecommandlockouts
	
\begin{document} 
	\maketitle
	\thispagestyle{empty}

	\begin{abstract}
	This paper presents a unified and scalable framework for predictive and safe autonomous navigation in dynamic transportation environments by integrating model predictive control (MPC) with distributed Koopman operator learning. High-dimensional sensory data are employed to model and forecast the motion of surrounding dynamic obstacles. A consensus-based distributed Koopman learning algorithm enables multiple computational agents or sensing units to collaboratively estimate the Koopman operator without centralized data aggregation, thereby supporting large-scale and communication-efficient learning across a networked system. The learned operator predicts future spatial densities of obstacles, which are subsequently represented through Gaussian mixture models. Their confidence ellipses are approximated by convex polytopes and embedded as linear constraints in the MPC formulation to guarantee safe and collision-free navigation. The proposed approach not only ensures obstacle avoidance but also scales efficiently with the number of sensing or computational nodes, aligning with cooperative perception principles in autonomous navigation applications. Theoretical convergence guarantees and predictive constraint formulations are established, and extensive simulations demonstrate reliable, safe, and computationally efficient navigation performance in complex environments.
	\end{abstract}
	

	\section{Introduction}
	\label{Sec. 1}
	The advent of autonomous navigation systems has driven increasing interest in autonomous agents that can safely navigate in dynamic, uncertain environments such as urban streets, plazas, and shared pedestrian zones \cite{farina2015application}. In these settings, safety and reliability become paramount~\cite{tordesillas2021faster}. Ground robots or autonomous platforms moving through cluttered, time-varying environments must continuously account for surrounding obstacles, adapt to uncertainty, and rapidly incorporate new sensory information~\cite{siegwart2011introduction}. Among control methodologies, model predictive control (MPC) stands out for autonomous navigation-related applications by optimizing trajectories while enforcing constraints such as collision avoidance and kinematic limits~\cite{rawlings2017model}.
	
	Avoiding collisions with moving targets such as vehicles, pedestrians, or crowds is a central concern in autonomous navigation applications. Considerable work has addressed static obstacles (see e.g.,~\cite{farina2015application, li2023survey}), but dynamic obstacle avoidance remains more challenging due to unpredictability, interaction effects, and environmental uncertainty~\cite{li2023moving, olcay2024dynamic, wei2022moving}. Indeed, those forecasts directly influence the control decisions in a receding-horizon framework. As a result, prediction-aware MPC formulations, where predicted obstacle regions are incorporated as time-varying constraints, have grown in interest in robotics and autonomous navigation communities~\cite{lindqvist2020nonlinear, zhu2023real, olcay2024dynamic}.
	
	In real-world autonomous navigation deployments, perception is often achieved through high-dimensional sensing modalities such as cameras, radar, or LiDAR units mounted on vehicles, roadside units, or drones~\cite{xu2025lv}. Extracting and modeling dynamic behavior of obstacles from such data is nontrivial, especially when explicit motion models are unknown or unreliable. A natural remedy is to treat the environment as a spatiotemporal density field e.g., occupancy, flow, or intensity maps, and to learn predictive models directly from raw sensor sequences. This shift toward data-driven forecasting enables better handling of complex, non-Newtonian motion patterns (e.g. crowds, group flows) as probabilistic density predictions~\cite{hafner2019learning}, which aligns well with autonomous navigation needs such as trajectory prediction, flow forecasting, and shared autonomous navigation.

    The Koopman operator theory originates from the seminal work of B.O. Koopman in the early 1930s~\cite{koopman1931hamiltonian}, where it was introduced as an infinite-dimensional linear operator acting on observables of nonlinear dynamical systems. Although initially developed within the context of Hamiltonian mechanics, its broader implications for nonlinear dynamics were recognized much later. Over the past two decades, the Koopman framework has experienced renewed interest, largely due to its systematic development and popularization, establishing its role as a unifying spectral theory for nonlinear systems and data-driven analysis~\cite{mezic2005spectral,mezic2013analysis}. For comprehensive historical background and mathematical foundations, we refer the reader to the expository overview in~\cite{mezic2021koopman}.
%
The Koopman operator framework forms a powerful backbone for such data-driven modeling. By lifting nonlinear dynamics into a function space, system evolution becomes linear in the space of observables~\cite{mezic2013analysis}. This enables use of linear-system tools for nonlinear tasks like prediction, estimation, and control. In practice, finite-dimensional approximations of the Koopman operator are built from time-series data and have shown success in system identification, control, and spatiotemporal forecasting~\cite{mezic2013analysis, bevanda2021koopman, otto2021koopman, williams2015data, proctor2016dynamic}. In autonomous navigation and robotics domains, Koopman-based methods have been applied to modeling~\cite{gu2023deep, tian2025physically}, planning~\cite{gutow2020koopman}, object tracking~\cite{comas2021self}, robust control~\cite{zhang2022robust}, and vehicular mobility systems~\cite{manzoor2023vehicular, zheng2024koopman}.

	Nonetheless, most Koopman-learning methods remain centralized~\cite{williams2015data,korda2018linear,brunton2016discovering}. Centralized strategies require full access to the entire dataset at one location, an assumption that fails under autonomous navigation constraints, such as limited computational capabilities, distributed edge sensors, privacy rules, and bandwidth bottlenecks. Distributed Koopman learning addresses these limitations by allowing agents to compute local operator estimates and coordinate to form a shared global Koopman model. In our framework, high-dimensional sensory snapshots (e.g. partitions of camera images) are processed in parallel across multiple agents. Each agent works on its share of the lifted observables and participates in a consensus-based least-squares scheme to estimate the global Koopman operator without exchanging raw data. This mechanism enhances scalability, reduces communication overhead, and ensures adaptability. The resulting operator can then propagate high-dimensional sensory states forward in time to generate spatial density forecasts of dynamic obstacles.
	
	Prior works have explored various distributed or scalable Koopman paradigms, yet important gaps persist in combining convergence, interpretability, and computational efficiency. The alternating minimization with neural liftings in~\cite{liu2020towards} introduces nonconvexity and dictionary-based splitting. The neighborhood-based graph Koopman in~\cite{nandanoori2021data} depends on known coupling structure and lacks a seamless least-squares consensus framework. The graph neural networks based geometric method in~\cite{mukherjee2022learning} trades off interpretability and incurs heavy communication overhead. Deep distributed Koopman methods (e.g.\ \cite{hao2024distributedAAA, hao2024distributedBBB}) enhance expressivity at the price of non-convexity and  large communication. The sequential partitioning scheme in~\cite{azarbahram2025distributed} handles temporal splits and ensures convergence to extended dynamic mode decomposition (EDMD), but fails to partition high-dimensional spatial data.
	
	We propose a distributed Koopman learning algorithm in which agents process local sensor data and collaboratively recover a global Koopman operator via consensus-based optimization over a communication graph. The formulation preserves the convexity and interpretability of EDMD and guarantees convergence to the centralized solution under mild connectivity and stepsize conditions, while avoiding nonconvex optimization, raw data exchange, and excessive communication. This enables scalable learning in high-dimensional lifted spaces, making the approach well suited for autonomous navigation scenarios such as imaging-based perception, urban crowd dynamics, and vehicular sensing. The proposed method complements temporal data-splitting approaches such as~\cite{azarbahram2025distributed} by addressing an orthogonal, spatially distributed data regime.  

The Koopman operator framework has recently gained traction for data-driven modeling and prediction in robotic and navigation systems. In~\cite{bueno2025koopman}, Koopman-based models were introduced for forecasting unknown moving obstacles in single unmanned aerial vehicle (UAV) navigation. This idea was extended to cooperative multi-UAV settings in~\cite{azarbahram2025distributedKoopman} through distributed and switched MPC for decentralized collision-free navigation. In contrast, the present work introduces a fully distributed Koopman learning architecture that learns the dynamic obstacle model itself without centralized fusion and integrates the learned operator into an MPC framework for safe navigation. Predicted obstacle densities are encoded via Gaussian mixtures and conservative polytopic approximations, which are embedded as linear constraints in MPC, enabling scalable and real-time deployment in large-scale, sensor-rich autonomous systems.

	The remainder of the paper is organized as follows.  The introduction continues with notations and graph definitions; Section~II presents the problem formulation and necessary preliminaries; Section~III describes the distributed Koopman algorithm and its integration with MPC; Section~IV presents simulation results validating the method; and Section~V concludes the paper.
	
	\medskip
	\paragraph*{Notations and definitions}
	Throughout the paper, $\N$, $\R$, and $\C$ denote the sets of nonnegative integers, real numbers, and complex numbers, respectively. Vectors in $\R^n$ and real matrices in $\R^{n\times m}$ are written using the customary superscripts. The $n\times n$ identity is $I_n$, while $0_{n\times m}$ and $1_{n\times m}$ are the all–zeros and all–ones matrices of size $n\times m$. When the size can be inferred, subscripts are omitted for brevity. Diagonal constructions use $\diag(\cdot)$: for scalars $a_1,\dots,a_p$ we write $\diag(a_1,\dots,a_p)$, and for blocks $A_1,\dots,A_p$ we use $\diag(A_1,\dots,A_p)$. The Kronecker product is $A\otimes B$, and matrix transpose is $A^\top$.  
	For a square matrix $A\in\R^{n\times n}$, we use $\det(A)$ for the determinant, $\trace(A)$ for the trace, and $\Lambda(A)$ for the spectrum (set of eigenvalues). If $\lambda\in\C$, its real and imaginary parts are denoted by $\re(\lambda)$ and $\im(\lambda)$.  
	For vectors $x\in\R^n$, the Euclidean norm is $\|x\|$. For matrices $A\in\R^{m\times n}$, the Frobenius norm is $\|A\|_F:=\trace(A^\top A)$. \\
	Let ${G}=(\cP,\cE)$ be an undirected graph with node set $\cP:=\{1,\dots,p\}$, $p\in\N$, and edge set $\cE\subseteq \cP\times\cP$. An edge $(i,j)\in\cE$ implies the reciprocal edge $(j,i)\in\cE$. A \emph{path} is a sequence of nodes linked by edges; the graph is \emph{connected} if every pair of nodes is joined by some path. For each node $i\in\cP$, its neighborhood is $N(i):=\{j\in\cP:(i,j)\in\cE\}$. The Laplacian matrix associated with graph $G$ is ${L}=[{L}_{ij}]\in\R^{p\times p}$, defined elementwise by
	\[
	{L}_{ij}=\begin{cases}
	-1, & \text{if } (i,j)\in\cE,\; i\neq j,\\[4pt]
	0, & \text{if } (i,j)\notin\cE,\; i\neq j,\\[4pt]
	-\displaystyle\sum_{k\neq i}{L}_{ik}, & \text{if } i=j.
	\end{cases}
	\]
	
	\section{Preliminaries and Problem Statement}
	\label{Sec. 2} 
	In this section, we first give some preliminaries and then the safe navigation control problem is formulated.
	
	\subsection{Unicycle robot model and feedback linearization}
	The unicycle robot model is adopted as a representative abstraction of wheeled autonomous ground vehicles widely studied in autonomous navigation communities. Its dynamics capture the essential nonholonomic motion constraints of cars, delivery robots, and small-scale urban service platforms, while remaining analytically tractable for control design and verification. This simplified model provides a canonical testbed for developing and validating predictive control strategies in dynamic traffic-like settings, where motion feasibility, safety margins, and real-time trajectory planning are critical. Beyond ground vehicles, the same modeling principles extend naturally to other nonholonomic platforms such as mobile robots and automated guided vehicles used in cooperative transportation and urban mobility networks.
	The considered kinematics model of the unicycle robot is captured by the nonlinear system
	\begin{equation}
	\left\{
	\begin{aligned}
	\dot{x} &= v \cos \theta, \\
	\dot{y} &= v \sin \theta, \\
	\dot{\theta} &= \omega, \\
	\dot{v} &= a,
	\end{aligned}
	\right.
	\label{eq:nonlinear_model}
	\end{equation}
	where the state \((x,y,\theta)\) encodes the global position and orientation, and the control variables are the linear acceleration \(a\) and angular velocity \(\omega\). The parameter \(v\) denotes the linear velocity.  

	To obtain a linear structure suitable for control, a feedback linearization scheme~\cite{oriolo2002wmr} is adopted. Introducing the transformed coordinates
	\[
	\eta_1 = x, \quad \eta_2 = \dot{x}, \quad \eta_3 = y, \quad \eta_4 = \dot{y},
	\]
	the dynamics in terms of \(\eta_i,\,i=1,2,3\) becomes
	\begin{subequations}
	\begin{align}
	\dot{\eta}_1 &= \eta_2, \\
	\dot{\eta}_2 &= a \cos \theta - v \omega \sin \theta, \\
	\dot{\eta}_3 &= \eta_4, \\
	\dot{\eta}_4 &= a \sin \theta + v \omega \cos \theta.
	\end{align}
	\end{subequations}
	
	\noindent
	By defining auxiliary control inputs
	\[
	a_x = a \cos \theta - v \omega \sin \theta, \qquad
	a_y = a \sin \theta + v \omega \cos \theta,
	\]
	the system is recast into two decoupled double integrators. The mapping back to the original inputs is recovered via
	\begin{equation}
	\begin{bmatrix}
	\omega \\
	a
	\end{bmatrix}
	= \frac{1}{v}
	\begin{bmatrix}
	- \sin \theta & \cos \theta \\
	v \cos \theta & v \sin \theta
	\end{bmatrix}
	\begin{bmatrix}
	a_x \\
	a_y
	\end{bmatrix},
	\label{eq:input_recovery}
	\end{equation}
	which is valid provided \(v \neq 0\). This singularity must be considered during controller implementation as emphasized in~\cite{oriolo2002wmr}.
	
	Let now the state of the linearized representation be \(\mathrm{x}_t = [\eta_1,\eta_2,\eta_3,\eta_4]^\top\). Its discrete-time evolution is modeled as
	\begin{equation}
	\mathrm{x}_{t+1} = A \mathrm{x}_t + B \mathrm{u}_t,
	\label{eq:linear_discrete_model}
	\end{equation}
	with input vector \(\mathrm{u}_t = [a_x,a_y]^\top\). The system matrices are
	\[
	A = \begin{bmatrix}
	1 & \tau & 0 & 0 \\
	0 & 1 & 0 & 0 \\
	0 & 0 & 1 & \tau \\
	0 & 0 & 0 & 1
	\end{bmatrix}, \quad
	B = \begin{bmatrix}
	\frac{\tau^2}{2} & 0 \\
	\tau & 0 \\
	0 & \frac{\tau^2}{2} \\
	0 & \tau
	\end{bmatrix},
	\]
	where \(\tau\) is the sampling interval.  
	
	The Cartesian position of the robot, required for reference tracking and obstacle avoidance, is obtained through
	\begin{equation}
	\mathrm{z}_t = C \mathrm{x}_t,
	\end{equation}
	where
	\[
	C = \begin{bmatrix}
	1 & 0 & 0 & 0 \\
	0 & 0 & 1 & 0
	\end{bmatrix}.
	\]
	This discretized and feedback-linearized formulation provides the predictive model used within the MPC framework described in the subsequent section.

\subsection{Koopman Operator Preliminaries}
    
	The Koopman operator framework provides a linear representation of nonlinear dynamical systems by acting on functions of the state rather than the state itself. Consider the discrete-time system
	\begin{align}
	x_{k+1} &= f(x_k), \quad x_k \in  \mathbb{R}^q, \label{eq:nonlinear_dynamics}
	\end{align}
	where the mapping \( f \) is nonlinear and \( q \) denotes the state dimension.  
	Instead of analyzing the state trajectory \( x_k \), one studies the evolution of the observable \(\Phi\) belonging to a Hilbert space \(\mathcal{H}\) of scalar-valued functions.  
	\textcolor{black}{The infinite-dimensional Koopman operator \(\mathcal{K}\) is introduced as 
	\begin{align}
	(\mathcal{K}\Phi)(x) = \Phi(f(x)), \quad \forall \Phi \in \mathcal{H}. \label{eq:koopman_def}
	\end{align}
	Although \(f\) is nonlinear, the operator \(\mathcal{K}\) acts linearly on the space of observables. To approximate its action in finite dimensions, we select \(n\) observables and group them into the vector-valued mapping $\bar{\bm{\Phi}}(x):\mathbb{R}^{q}\to\mathbb{R}^{n}$  
	\begin{align}
	\bar{\bm{\Phi}}(x) &= [\Phi_1(x), \Phi_2(x), \dots, \Phi_n(x)]^\top, \label{eq:observable_vector}
	\end{align}
	which lifts each state \(x \) into an \(n\)-dimensional feature space~\cite{williams2015data}.}
	With a dataset of \(N\) samples \(\{x_k, f(x_k)\}_{k=1}^N\), the lifted snapshots are organized as
	\begin{align}
	X &= [\bar{\bm{\Phi}}(x_1), \bar{\bm{\Phi}}(x_2), \dots, \bar{\bm{\Phi}}(x_N)] \in \mathbb{R}^{n \times N}, \label{eq:data_X} \\
	Y &= [\bar{\bm{\Phi}}(f(x_1)), \bar{\bm{\Phi}}(f(x_2)), \dots, \bar{\bm{\Phi}}(f(x_N))] \in \mathbb{R}^{n \times N}. \label{eq:data_Y}
	\end{align}
	The EDMD~\cite{williams2015data} provides a tractable approximation by projecting the action of \(\mathcal{K}\) onto the span of the chosen observables. This yields a finite matrix \(K \in \mathbb{R}^{n \times n}\) such that the lifted states evolve approximately according to
	\begin{align}
	Y &\approx K X. \label{eq:koopman_matrix_approx}
	\end{align}
	The optimal approximation is obtained from the Frobenius-norm least-squares problem
	\begin{align}
	K^* &= \arg\min_{K \in \mathbb{R}^{n \times n}} \left\| Y - K X \right\|_F^2, \label{eq:frobenius_minimization}
	\end{align}
	which provides a finite-dimensional representation of the Koopman operator consistent with the available data.

	\subsection{Problem statement}
We address safe and predictive navigation in dynamic environments by integrating distributed perception with MPC. High-dimensional spatiotemporal observations of moving obstacles, obtained from aerial camera snapshots, are modeled using a Koopman operator that lifts nonlinear obstacle dynamics into a linear predictive representation. To handle the data dimensionality, Koopman learning is performed in a distributed manner across multiple computational nodes via consensus, yielding a globally consistent operator without centralized processing. The resulting predictions are used to forecast obstacle density maps, which are approximated by Gaussian mixtures and converted into convex polytopic regions. These regions enter an MPC formulation for a unicycle robot as time-varying linear constraints, enabling anticipatory and collision-free motion planning in complex dynamic environments.

	\section{Main Results}
	 \label{Sec. 3} 
	
	In this section the proposed perception-to-control architecture is introduced.
	Given the goal position \( \mathrm{z}^* \) and the prediction horizon \( H \), 
	the task of the MPC controller is to ensure that the actual state and control input of the robot remain close to their reference counterparts while fulfilling all dynamic, actuation, and safety constraints derived from the environment perception model. 
	The optimization problem is formally stated as follows:
	\begin{subequations}\label{eq:robust_mpc}
	\begin{align}
	\min_{\bar{\mathrm{z}}_{t},\, \bar{\mathrm{u}}_{t}}  & 
	\sum_{h=0}^{H-1} \left\| {\mathrm{z}}_{t+h|t} - \mathrm{z}^* \right\|_Q^2 
	+ \left\| {\mathrm{u}}_{t+h|t} \right\|_R^2 
	 \label{eq:mpc_cost} \\
	\text{s.t.} \quad 
	& \left\{
	\begin{aligned}
	\mathrm{x}_{t+h+1|t} &= A \mathrm{x}_{t+h|t} + B {\mathrm{u}}_{t+h|t}, \\
	\mathrm{z}_{t+h|t} &= C \mathrm{x}_{t+h|t}, \quad \forall h = 0,\dots,H-1,
	\end{aligned}
	\right.  \label{eq:mpc_dyn} \\
	& \mathrm{x}_{t|t} = \mathrm{x}_t, \label{eq:mpc_ini} \\
	&\Omega_{t+h,\ell|t}^* ~  {\mathrm{z}}_{t+h|t}\geq \varrho_{t+h,\ell|t}^* + \varepsilon, \nonumber \\
	    &\qquad \forall h = 0, \dots, H-1, \quad \forall \ell = 1, \dots, L_{o}, \label{OsbtacleMPC}
	\end{align}
	\end{subequations}
	where $(\bar{\mathrm{z}}_{t},\, \bar{\mathrm{u}}_{t}) =: (\mathrm{z}_{t:t+H-1|t},\, {\mathrm{u}}_{t:t+H-1|t})$. 
	The symmetric positive matrices \( Q \) and \( R  \) include weighting coefficients that balance tracking accuracy against control effort, ensuring smooth and stable behavior. The initial condition of the optimization problem is set in~\eqref{eq:mpc_ini} and the constraints in~\eqref{OsbtacleMPC} are incorporated to guarantee obstacle-free trajectories by restricting the predicted robot positions within admissible safe regions. 
	While the specific structure of the parameters $\Omega_{t+h,\ell|t}^*$, $\varrho_{t+h,\ell|t}^*$, $L_{o}$, and $\varepsilon$ will be rigorously defined in the subsequent sections, it is worth noting that these terms are derived directly from distributively learned Koopman-based predictions of dynamic obstacle motion. Each constraint represents a linearized safety boundary that evolves in time based on the learned dynamics, thus transforming high-dimensional sensory forecasts into computationally efficient linear inequalities. 
	This tight integration between perception and control ensures that the MPC simultaneously reasons about goal tracking and future obstacle motion, setting the foundation for a predictive, adaptive navigation strategy.  
	
	The optimal control sequence obtained from solving~\eqref{eq:robust_mpc}, denoted by $(\bar{\mathrm{z}}_{t}^*,\, \bar{\mathrm{u}}_{t}^*)$, defines the desired trajectories over the prediction horizon, while only the first control input ${\mathrm{u}}_{t|t}^*$ is executed at each time step according to the receding horizon principle.
	
	\subsection{Distributed Koopman modeling of dynamic obstacles}

To enable safe and predictive navigation in dynamic environments, we model the motion of unknown obstacles without assuming explicit dynamics. A Koopman operator framework is adopted to lift nonlinear obstacle motion into a linear representation using high-dimensional perceptual data, such as image or video snapshots, which encode the spatiotemporal evolution of the environment.

This setting naturally motivates a distributed learning architecture in two practical scenarios. In the first, a single UAV observes the entire environment through a high-resolution sensor, but the resulting data dimensionality renders centralized Koopman learning computationally infeasible; the learning task is therefore distributed across onboard computational cores, which act as virtual agents processing subsets of the lifted observables. In the second, a team of UAVs with limited, possibly overlapping fields of view collects partial observations of the environment; here, the agents correspond to physical UAVs that communicate over a network to collaboratively learn a unified Koopman model. In both cases, the common challenge is to learn a global Koopman operator from partitioned, high-dimensional data under communication and computational constraints.

Accordingly, each agent \( i \in \{1,\dots,p\} \) has access only to a local segment of the raw one-step data,
\begin{align*}
    \mathcal{X}_i=\begin{bmatrix}\bar x_{i,1}&\cdots&\bar x_{i,N}\end{bmatrix}\in\mathbb{R}^{q_i\times N}, \nonumber \\
\mathcal{Y}_i=\begin{bmatrix}\bar y_{i,1}&\cdots&\bar y_{i,N}\end{bmatrix}\in\mathbb{R}^{q_i\times N},
\end{align*}
where $\bar y_{i,k}$ denotes the forward transition of $\bar x_{i,k}$ and $q_i$ is the local data dimension. Local lifting maps $\bar{\bm{\Phi}}_i:\mathbb{R}^{q_i}\to\mathbb{R}^{n_i}$ generate the lifted data
\[
X_i=\bar{\bm{\Phi}}_i(\mathcal{X}_i)\in\mathbb{R}^{n_i\times N}, \quad
Y_i=\bar{\bm{\Phi}}_i(\mathcal{Y}_i)\in\mathbb{R}^{n_i\times N},
\]
with $\sum_{i=1}^{p} n_i = n$, yielding the global structure
\[
X=\begin{bmatrix}X_1\\ \vdots\\ X_p\end{bmatrix}, \quad
Y=\begin{bmatrix}Y_1\\ \vdots\\ Y_p\end{bmatrix},
\]
consistent with \eqref{eq:data_X}–\eqref{eq:data_Y}. Each agent aims to compute its block-column \( K_i\in\mathbb{R}^{n\times n_i} \) of the global Koopman operator by collaboratively solving the least-squares problem~\eqref{eq:frobenius_minimization} through distributed optimization and consensus over a connected undirected communication graph \( {G} \), without central data aggregation.

	The group of agents aims to solve the problem \eqref{eq:frobenius_minimization} by a distributed algorithm as
	\begin{subequations}\label{algo2}
	\begin{align}
		K_i^+&=K_i-\alpha S_i X_i^\top,\label{algo2_1}\\
		S_i^+&=S_i+(K_i^+-K_i)X_i-\alpha\sum_{j\in N(i)}(S_i-S_j).\label{algo2_2}
	\end{align}
	\end{subequations}
	Here, $\alpha>0$ is a discretization step-size and $S_i\in\R^{n\times N}$ is another internal augmented state variable owned by the $i$-th agent for facilitating the convergence. The variables $K_i, S_i$ start from the specific initial conditions
	\begin{equation}\label{initial_condition}
	K_i(0)=0, \quad S_i(0)=-\bm Y_i\quad\forall i\in\cP,
	\end{equation}
	where $\bm Y_i\in\R^{n\times N}$ is the $i$-th block-column of the block-diagonal matrix
	\begin{equation}\label{def:bA}
	\bm Y:=\diag(Y_1,\cdots,Y_p)\in\R^{n\times Np}.
	\end{equation}
	
	\begin{algorithm}[t]
	\caption{Distributed Koopman Operator Learning}\label{algo}
	\begin{algorithmic}[1]
	\Require $\alpha$ satisfying $\alpha<\alpha_{\max}$ from \eqref{def:alpha_max}. 
	\State Initialize $K_i(0)\in\R^{n\times n_i}$, $S_i(0) \in\R^{n\times N}$ for all $i\in\cP$ according to \eqref{initial_condition}.
	\Loop{ for $t=0,1,\ldots,t_{\max}-1$, the $i$-th agent, $i\in\cP$}
	\State Broadcast $S_i(t)$ to its neighbors.
	\State Compute $K_i(t+1), S_i(t+1)$ according to \eqref{algo2}.
	\EndLoop
	\Ensure $K(t_{\max})=\begin{bmatrix}
			K_1(t_{\max})&\cdots&K_p(t_{\max})
		\end{bmatrix}$.
	\end{algorithmic}
	\end{algorithm}
	
	The convergence of the algorithm \eqref{algo2} is related to the eigenvalues of the following matrix
	\begin{equation}\label{def:bM}
	\bM:=-\begin{bmatrix}
		\bm X \bm X^\top & \bm X \bL\\\bm X^\top & \bL
	\end{bmatrix}\in \R^{(Np+n)\times (Np+n)},
	\end{equation}
	where
	\begin{align}
	\bm X &=\diag(X_1,\cdots,X_p)\in\R^{n\times Np},\label{def:bB}\\
	\bL&=L\otimes I_N\in\R^{Np\times Np},\label{def:bL}
	\end{align}
	and $L$ is the Laplacian matrix of ${G}$.  The distributed algorithm for Koopman operator learning is summarized in Algorithm~\ref{algo}.
	    Before introducing the main result in Theorem~\ref{thm:row}, the following lemma is instrumental. 
	\begin{lemma}\label{lem:row}
	$K^*\in \R^{n\times n}$ is an optimal solution to the problem \eqref{eq:frobenius_minimization} if and only if there exists $W^*\in\R^{n\times Np}$ such that $(K^*, W^*)$ is an optimal solution to the constrained problem
	\begin{subequations}\label{op_row_compact}
		\begin{align}
			\min_{K,W}&\Vert \bm Y-K\bm X-W\Vert^2,\\
			&\mbox{subject to } W (1_p\otimes I_N)=0,
		\end{align}		
	\end{subequations}
	where $\bm Y,\bm X$ are defined in \eqref{def:bA}, \eqref{def:bB}.
	\end{lemma}
	
	\begin{proof}
	Define the Lagrangian 
	\begin{equation*}
		\cL(K,W,\Lambda)=\frac{1}{2}\Vert \bm Y-K\bm X-W\Vert^2+ \tr(\Lambda^\top W (1_p\otimes I_N)),
	\end{equation*}
	where $\Lambda\in\R^{n\times Np}$ is the matrix of Lagrangian multipliers.	
	Note that the optimization problem \eqref{op_row_compact} is convex; hence by the KKT condition \cite{Boyd_Vandenberghe_2004}, $(K^*, W^*)\in \R^{n\times n}\times \R^{n\times Np }$ is an optimal solution to \eqref{op_row_compact} if and only if for some $\Lambda^*\in \R^{n\times Np }$,
	\begin{align}
		\frac{\partial L}{\partial K}(K^*,W^*,\Lambda^*)=&-(\bm Y-K^*\bm X-W^*)\bm X^\top =0,\label{KKT1}\\
		\frac{\partial L}{\partial W}(K^*,W^*,\Lambda^*)=&-(\bm Y-K^*\bm X-W^*)\nonumber\\
		&\quad+\Lambda^*(1_p^\top \otimes I_N) =0,\label{KKT2}\\
		\frac{\partial L}{\partial \Lambda}(K^*,W^*,\Lambda^*)=&W^* (1_p\otimes I_N)=0.\label{KKT3}
	\end{align}
	From \eqref{KKT2}, 
	\begin{equation}\label{W^*}
		W^*=\bm Y-K^*\bm X-\Lambda^*(1_p^\top \otimes I_N).
	\end{equation}
	Substituting into \eqref{KKT3}, we obtain
	\begin{equation*}
		(\bm Y-K^*\bm X)(1_p\otimes I_N)=\Lambda^*(1_p^\top \otimes I_N)(1_p\otimes I_N)=p\Lambda^*.
	\end{equation*}
	Therefore, 
	        $$\Lambda^*=\frac{1}{p}(\bm Y-K^*\bm X)(1_p\otimes I_N)$$
	        which, after substituting into \eqref{W^*}, gives
	\begin{equation*}
		W^*=(\bm Y-K^*\bm X)\left(I_{Np}-\frac{1}{p}(1_p1_p^\top)\otimes I_N\right).
	\end{equation*}
	Substituting into \eqref{KKT1}, we conclude that
	\begin{equation}\label{orthogonality}
		\frac{1}{p}(\bm Y-K^*\bm X) \left((1_p1_p^\top)\otimes I_N\right)\bm X^\top =0.
	\end{equation}
	In other words, $K^*$ needs to be such that $(\bm Y-K^*\bm X)$ is orthogonal to $\bm X\left((1_p1_p^\top)\otimes I_N\right)$.
	
	On the other hand, note that the optimization problem \eqref{eq:frobenius_minimization} can be equivalently written as
	\begin{equation}
		\min_{K}\Vert(\bm Y-K\bm X)(1_p\otimes I_n)\Vert^2,
	\end{equation}
	which is also convex. Hence, the first order condition, 
	\begin{equation}\label{first_order}
		(\bm Y-K^*\bm X)(1_p\otimes I_n)(1_p\otimes I_n)\bm X^\top=0,
	\end{equation}
	is both necessary and sufficient for optimality. Because \eqref{first_order} is the same as \eqref{orthogonality}, Lemma~\ref{lem:row} is proven.
	\end{proof}
	    
	    We are now in a position to introduce the following theorem.
	\begin{thm}\label{thm:row}
	Suppose the communication graph ${G}$ is undirect and connected. There exists
	\begin{equation}\label{def:alpha_max}
		\alpha_{\max}:=-\max_{\lambda\in\Lambda(\bM)\backslash\{0\}}\frac{2\re(\lambda)}{|\lambda|^2}>0,
	\end{equation}
	such that as long as the step size $\alpha<\alpha_{\max}$, the matrix
	\begin{equation}
		K=\begin{bmatrix}
			K_1&\cdots&K_p
		\end{bmatrix}
	\end{equation}
	given by \eqref{algo2} starting from initial conditions \eqref{initial_condition} will converge to an optimal solution for the problem~\eqref{eq:frobenius_minimization}. Furthermore, the convergence is exponential with rate $\rho>\rho_{\max}$, where
	\begin{equation}\label{def:rho_max}
		\rho_{\max}:=\max_{\lambda\in\Lambda(\bM)\backslash\{0\}}\sqrt{1+2\alpha\re(\lambda)+\alpha^2|\lambda|^2}.
	\end{equation}
	\end{thm}
	\begin{proof}
	Define
	\begin{align*}
		W&=S+\bm Y-K \bm X,\\
		e_K&=K-K^*, \quad e_W=W-W^*.
	\end{align*}
	It follows from \eqref{algo2_1} and \eqref{KKT1} that
	\begin{align*}
		e_K^+&=K^+-K^*\\
		&=K-\alpha S\bm X^\top-K^*\\
		&=e_K+\alpha (\bm Y-K \bm X-W)\bm X^\top\\
		&=e_K+\alpha (\bm Y-(e_K+K^*) \bm X-(e_W+W^*))\bm X^\top\\
		&=e_K-\alpha(e_K\bm X +e_W)\bm X^\top.
	\end{align*}
	Meanwhile, Because the graph ${G}$ is connected, $1_p^\top L=0$. Thus \eqref{KKT2} implies $(\bm Y-K^*\bm X-W^*)\bL=0$, where recall $\bL$ is defined by \eqref{def:bL}. Hence it follows from \eqref{algo2_2} that
	\begin{align*}
		e_W^+&=W^+-W^*\\
		&=S^++\bm Y-K^+\bm X-W^*\\
		&=S+(K^+-K)\bm X+\alpha (\bm Y-K\bm X-W)\bL\\
		&\quad+\bm Y-K^+\bm X-W^*\\
		&=S+\bm Y-K\bm X+\alpha (\bm Y-K\bm X-W)\bL-W^*\\
		&=W+\alpha (\bm Y-K\bm X-W)\bL-W^*\\
		&=e_W+\alpha (\bm Y-(e_K+K^*)\bm X-(e_W+W^*))\bL\\
		&=e_W-\alpha(e_K\bm X+e_W) \bL. 
	\end{align*}
	Because of the initial conditions \eqref{initial_condition}, $W(0)=0$, which implies $e_W(0)=-W^*$. Therefore, $e_W(k)(1_p\otimes I_N)=0$ for all $k\in\N$. 	
	In summary, we have
	\begin{equation}
		\begin{bmatrix}
			e_K^+&e_W^+
		\end{bmatrix}=\begin{bmatrix}
			e_K&e_W
		\end{bmatrix}\left(I+\alpha\bM\right),
	\end{equation}
	where recall $\bM$ is defined in \eqref{def:bM}.
	It can be verified that $\bM$ is semi-Hurwitz. Hence when $\alpha<\alpha_{\max}$, where $\alpha_{\max}$ is defined in \eqref{def:alpha_max}, $I+\alpha\bM$ is semi-Schur. Consequently, by \cite[Corollary 3]{SL:24-arXiv}, $\begin{bmatrix}
		e_K&e_W
	\end{bmatrix}\to \begin{bmatrix}
		e_K^\infty&e_W^\infty
	\end{bmatrix}$ exponentially fast, such that $\begin{bmatrix}
		e_K^\infty&e_W^\infty
	\end{bmatrix}\bM=0$. This implies that
	\begin{align*}
		(e_K^\infty\bm X+e_W^\infty)\bm X^\top&=0,\\
		(e_K^\infty\bm X+e_W^\infty)\bL&=0.
	\end{align*}
	Making reference to \eqref{KKT1}-\eqref{KKT3}, we see that the limit
	\begin{equation*}
		\lim_{k\to\infty}\begin{bmatrix}
			K(k)&W(k)
		\end{bmatrix}=\begin{bmatrix}
			K^*&W^*
		\end{bmatrix}+\begin{bmatrix}
			e_K^\infty&e_W^\infty
		\end{bmatrix}
	\end{equation*}
	also satisfies the KKT condition for the problem \eqref{op_row_compact}. Hence by Lemma~\ref{lem:row}, $\lim_{k\to\infty}K(k)$ is an optimal solution to the problem~\eqref{eq:frobenius_minimization} and the convergence is exponentially fast.
	\end{proof}

    \begin{remak} \label{remark__alphamax}
The stepsize threshold $\alpha_{\max}$ in Theorem~\ref{thm:row} is expressed in terms of the eigenvalues of the matrix $\bM$, which depends on the global lifted data through $\bm X=\diag(X_1,\dots,X_p)$ and on the communication graph. As a result, computing $\alpha_{\max}$ exactly requires centralized spectral information and is not directly feasible in a distributed setting. Importantly, the distributed algorithm does not require the exact value of $\alpha_{\max}$: convergence is guaranteed for any stepsize $\alpha$ satisfying $\alpha<\alpha_{\max}$. In practice, such a stepsize can be selected conservatively using distributedly computable bounds. By combining local evaluations of $\|X_i\|_2$ (aggregated via max-consensus to obtain $\|\bm X\|_2=\max_i\|X_i\|_2$) with an upper bound on $\lambda_{\max}(L)$. This yields a common stepsize shared by all agents and ensures convergence of Algorithm~\ref{algo} without centralized coordination.
\end{remak}
	
	\subsection{Obstacle prediction and constraint formulation for MPC}
	
	The distributed Koopman learning algorithm introduced earlier serves as the foundation for prediction-based obstacle avoidance. 
	To clarify the Koopman prediction process, the paired matrices \( X \) and \( Y \)  encode the temporal transitions of the spatial density field observed by the robot or aerial sensor. Once the distributed Koopman operator \( K \) is estimated, it serves as a linear propagator:
	\[
	X_{t+h} \approx K^h X_t, \quad h = 1, 2, \dots, H.
	\]
	Repeated application of \( K \) to the latest lifted snapshot yields a sequence of predicted lifted states, which are mapped back to the spatial domain to reconstruct predicted density maps \( \rho_{t+h} \), \( h = 1, 2, \dots, H \). These maps represent the expected evolution of the dynamic obstacles over the MPC horizon. This procedure effectively translates temporal patterns captured from prior data into forward predictions of spatial occupancy, forming the basis for subsequent Gaussian mixture modeling and polytope-based constraint formulation.
	
	Therefore, using the distributed Koopman operator \( K \), we propagate lifted observables forward, reconstructing a sequence of density maps \(\{\rho_t\}_{t=1}^H\) over a 2D grid. Each \(\rho_t\) encodes predicted obstacle distributions. Thresholding \(\rho_t\) at \(c_\rho\) yields:
	\[
	\Psi_t = \left\{ \xi \in \mathbb{R}^2 : \rho_t(\xi) > c_{\rho} \right\}.
	\]
	We fit a \(L_{o}\)-component Gaussian mixture model to \(\Psi_t\), where each component \(\mathcal{N}(\xi \mid \mu_\ell, \Sigma_\ell)\) models a local obstacle as an ellipse.
	
	Each Gaussian mixture model component defines an elliptical region:
	\[
	\mathscr{C}_\ell = \left\{ \xi : (\xi - \mu_\ell)^\top \Sigma_\ell^{-1} (\xi - \mu_\ell) \leq \delta^2 \right\},
	\]
	with \(\delta^2\) set by a chi-squared quantile. Decomposing \(\Sigma_\ell = U D U^\top\), the semi-axes are:
	\[
	\breve{a}_\ell = \sqrt{\delta^2 \lambda_1}, \quad \breve{b}_\ell = \sqrt{\delta^2 \lambda_2}.
	\]
	Together with rotation \(U\), these parameters fully describe \(\mathscr{C}_\ell\).
	To form linear MPC constraints, we approximate \(\mathscr{C}_\ell\) by a polytope with \({n_{\varsigma}}\) half-spaces (supporting hyperplanes).
	
	Koopman-based forecasts produce time-indexed spatial density maps that are compressed into polytopes via confidence contours. Each polytope acts as a moving keep-out region; enforcing its most active supporting halfspace at each step is equivalent to shifting the robot's admissible output set away from predicted obstacles with a tunable margin. This converts perception into linear time-varying constraints that the MPC can handle natively, aligning the robot's feasible region with the anticipated flow of crowds. As a result, prediction and planning are tightly coupled: improved forecasts directly enlarge the feasible corridor, while conservative bounds still guarantee safety without destabilizing the optimizer.
	
	 For each angle $\vartheta_\iota = 2\pi(\iota-1)/{n_{\varsigma}},~ \iota = 1,\ldots,{n_{\varsigma}}$, we compute support points and normals:
	\[
	s_{\iota} = \mu_\ell + U
	\begin{bmatrix}
	  \breve{a}_\ell \cos \vartheta_{\iota} \\
	  \breve{b}_\ell \sin \vartheta_{\iota}
	\end{bmatrix}, \quad
	n_{\iota} = \begin{bmatrix} \cos \vartheta_{\iota} \\ \sin \vartheta_{\iota} \end{bmatrix}.
	\]
	Each constraint becomes \(n_{\iota}^\top \xi \geq n_{\iota}^\top s_{\iota} + \varepsilon\), and stacking gives:
	\[
	\Omega_\ell \xi \geq \varrho_\ell + \varepsilon \mathbf{1}_{n_{\varsigma}},
	\]
	with \(\Omega_\ell\) and \(\varrho_\ell\) collecting all normals and support values.
	%
	%
	At each MPC step \(t+h\), for \(h = 1, \dots, H\), and each obstacle \(\ell = 1, \dots, L_{o}\), we enforce:
	\[
	\Omega_{t+h,\ell} \xi_{t+h} \geq \varrho_{t+h,\ell} + \varepsilon \mathbf{1}_{n_{\varsigma}}.
	\]
	To reduce complexity, we identify and enforce only the most active facet
	\[
	\iota^* = \arg \max_{\iota} \left( {\Omega}_{t+h,\ell}^{(\iota)\top} \xi_{t+h} - \varrho_{t+h,\ell}^{(\iota)} - \varepsilon \right),
	\]
	yielding:
	\[
	\Omega_{t+h,\ell}^* := {\Omega}_{t+h,\ell}^{(\iota^*)\top}, \quad \varrho_{t+h,\ell}^* := \varrho_{t+h,\ell}^{(\iota^*)}.
	\]
	These constraints are reformulated in terms of the MPC output \(\mathrm{z}_{t+h|t}\), resulting in the final avoidance condition:
	\begin{align} \label{OBSTACLEFINAL}
	   \Omega_{t+h,\ell|t}^* ~  &{\mathrm{z}}_{t+h|t}\geq \varrho_{t+h,\ell|t}^* + \varepsilon, \nonumber \\
	    &\qquad \forall h = 0, \dots, H-1, \quad \forall \ell = 1, \dots, L_{o}.
	\end{align}
	The safety margin \(\varepsilon \geq R_r\) accounts for the robot’s size, ensuring robust obstacle avoidance.

	\section{Simulation Case-Study}
\label{Sec. 4}

\begin{figure}[!t]
	    \centering
	    \includegraphics[trim=0.0cm 0.0cm 11.0cm 5.0cm, clip, width=0.45\textwidth]{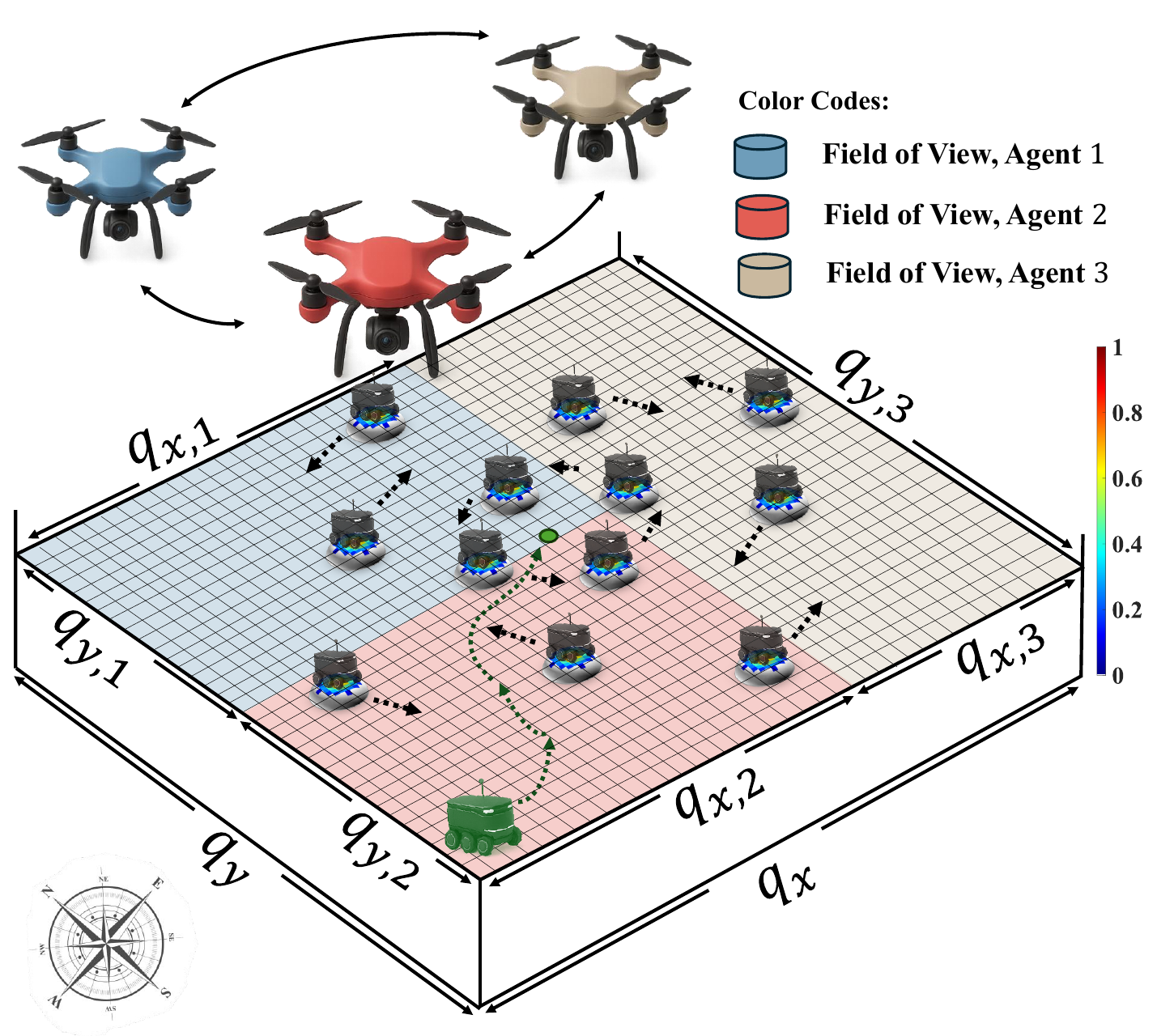}
	    \captionsetup{font=normalsize}
        \caption{Graphical illustration of the distributed perception-learning-control scenario. }
	    \label{Graphical_Example}
	\end{figure}

 A graphical overview of the simulated perception-learning-control scenario is provided in Fig.~\ref{Graphical_Example}. The setup consists of $p=3$ UAV agents collaboratively monitoring a planar area of dimension $q=q_x q_y$, corresponding to the state space of the unknown nonlinear dynamics in~\eqref{eq:nonlinear_dynamics}. The UAVs are connected through a circular undirected communication graph and are color-coded for clarity. Each UAV observes only a local field of view, indicated by the highlighted ground grids with matching colors. In particular, the field of view of agent~$i$ has dimension $q_i=q_{x,i}q_{y,i}$; in the simulations we set $q_x=q_y=30$, yielding $q=900$, and choose an equal spatial partition with $q_1=q_2=q_3=300$.

The monitored area contains multiple moving robot-like obstacles, depicted in black, each following the direction indicated by dashed arrows. The sensory snapshots collected by the UAVs are represented as two-dimensional occupancy grids with values in $[0,1]$, where values close to $1$ indicate dense obstacle regions and values near $0$ correspond to free space. Sequential snapshots collected over time form a spatiotemporal dataset capturing the evolution of these dynamic obstacles. In the simulations, each UAV records local snapshots over $N=10$ time steps using onboard camera measurements; the corresponding intensity color bar is shown in Fig.~\ref{Graphical_Example}.  

To enable distributed Koopman learning, local grid vectorization is employed as the lifting function, so that $n_i=q_i$ for all agents. In parallel, a nonholonomic ground robot (shown in green) navigates the environment toward a target location indicated by a green circle at the center of the grid. The dotted green arrows illustrate candidate trajectories generated by the MPC scheme. Predictions obtained from the distributed learning of obstacle motion are used to forecast future obstacle density maps, which are approximated by Gaussian mixture models and converted into convex polytopic regions. These polytopes enter the MPC formulation as time-varying linear constraints, enabling anticipatory and collision-free motion planning in a dynamic environment.

 \begin{figure}[b]
	    \centering
    \includegraphics[trim=0.0cm 0.0cm 0.0cm 0.0cm, clip, width=0.45\textwidth]{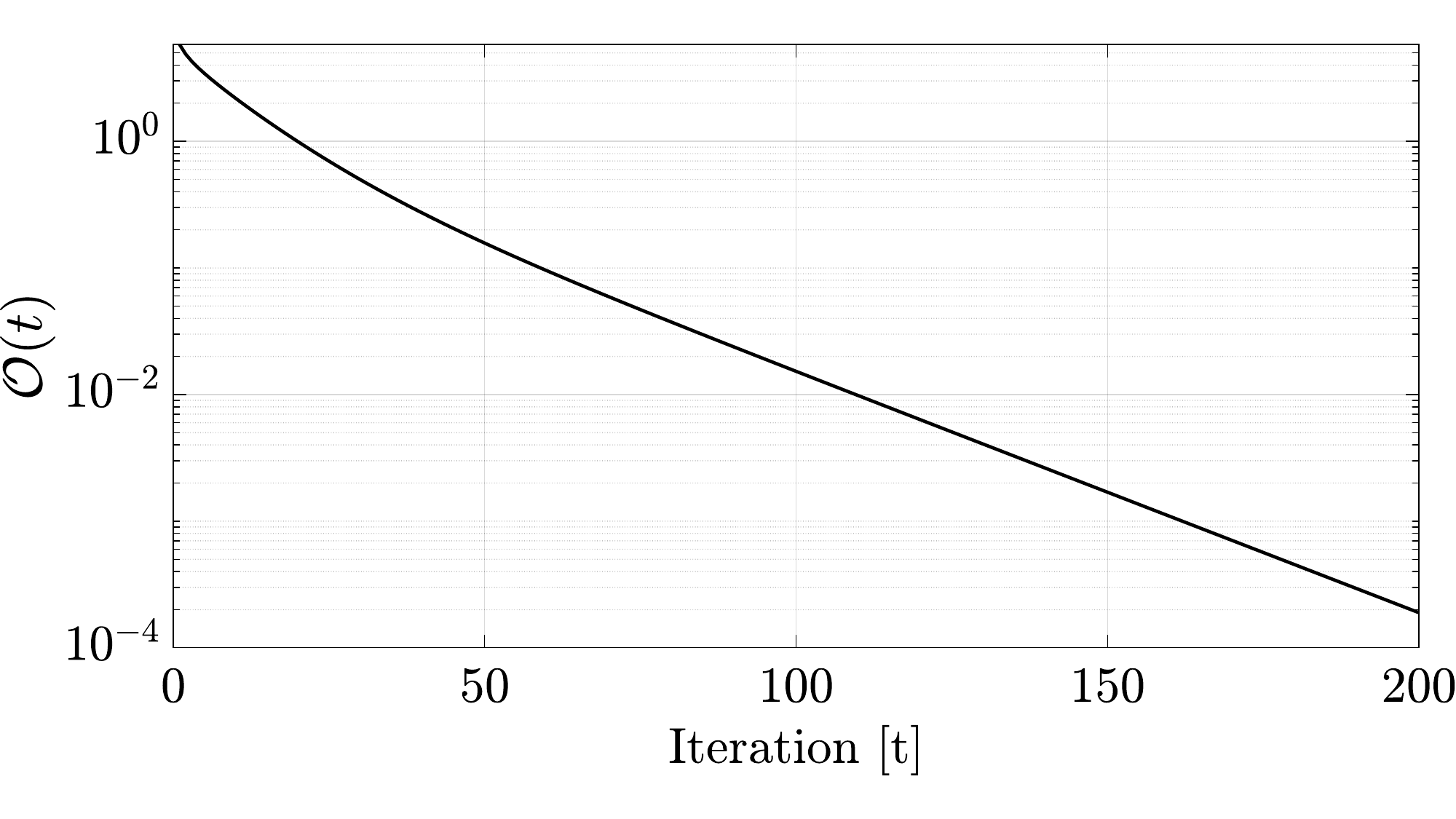}
	\captionsetup{font=normalsize}
	    \caption{Prediction discrepancy \( \mathcal{O}(t) \) between distributed and centralized Koopman operators over iterations.}
	    \label{fig:prediction_discrepancy}
	\end{figure}

\subsection{Evaluation of distributed Koopman operator learning}

The distributed Koopman learning algorithm is executed for $200$ optimization steps with neighbor communication. Using the matrix $\bM$ defined in~\eqref{def:bM}, we compute $\alpha_{\max}=0.2095$ and select the stepsize as $\alpha=0.5\,\alpha_{\max}$, yielding a contraction factor $\rho_{\max}=0.9185$ according to~\eqref{def:rho_max}. As discussed in Remark~\ref{remark__alphamax}, this step-size can be chosen in a distributed manner based on global bounds obtained via local computations and consensus, and can be shared consistently by all agents throughout the learning process.

\begin{figure}[!t]
	    \centering
        \includegraphics[trim=3.0cm 0.0cm 3.0cm 0.0cm, clip, width=0.45\textwidth]{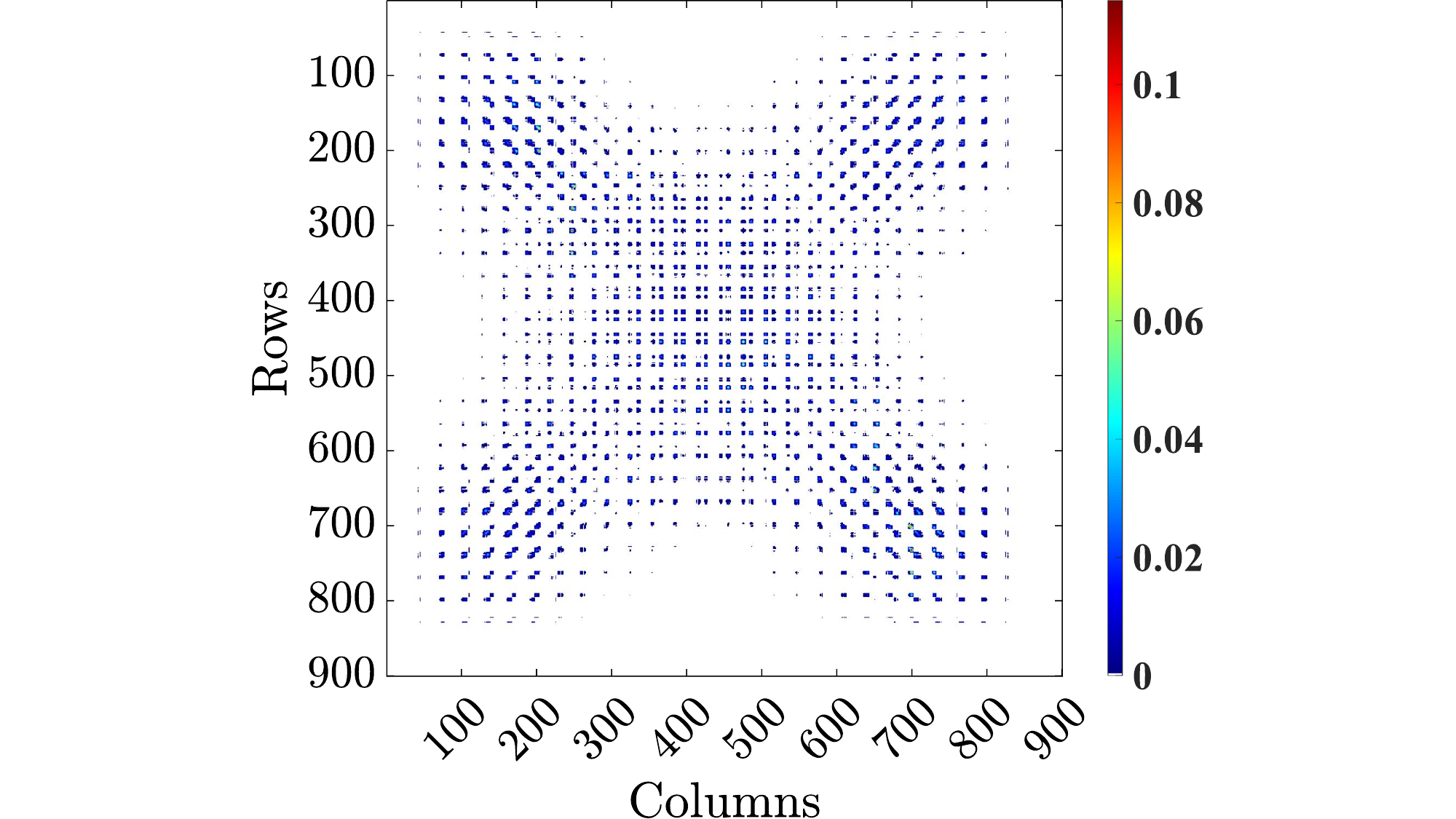}
	\captionsetup{font=normalsize}
	    \caption{Elementwise absolute difference between the distributed Koopman operator \( K_\mathrm{d} \) and the centralized Koopman operator \( K^* \).}
	 \label{fig:K_diff}
	\end{figure}

\begin{figure}[!t]
	    \centering
	     \includegraphics[trim=3.0cm 0.0cm 3.0cm 0.0cm, clip, width=0.45\textwidth]{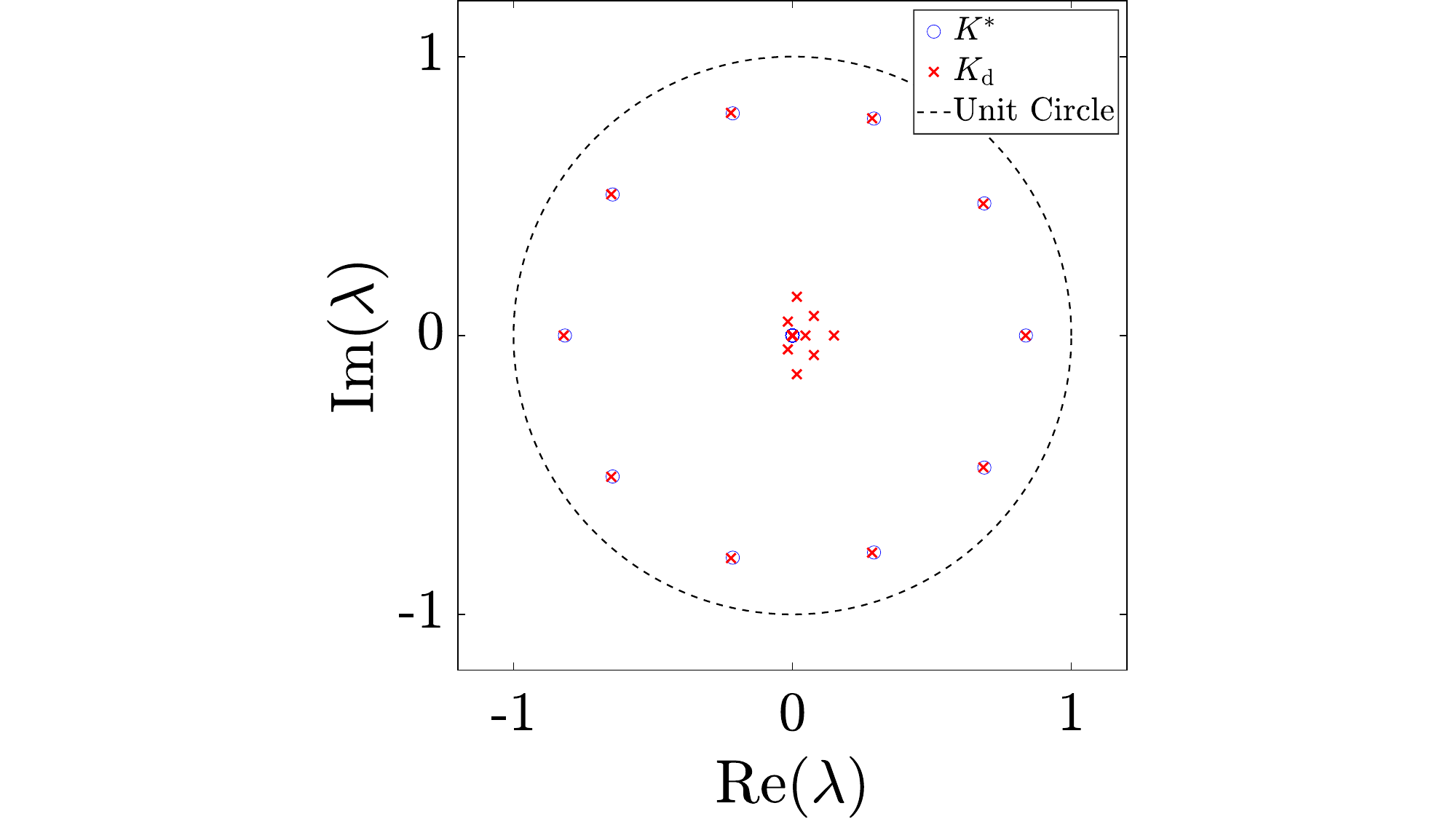}
	\captionsetup{font=normalsize}
	    \caption{Spectral comparison between the distributed Koopman operator \( K_\mathrm{d} \) and the centralized Koopman operator \( K^* \). }
	   \label{fig:eigenvalues}
	\end{figure}

\begin{figure}[!b]
	    \centering
	    \includegraphics[trim=0.0cm 0.0cm 0.0cm 0.0cm, clip, width=0.45\textwidth]{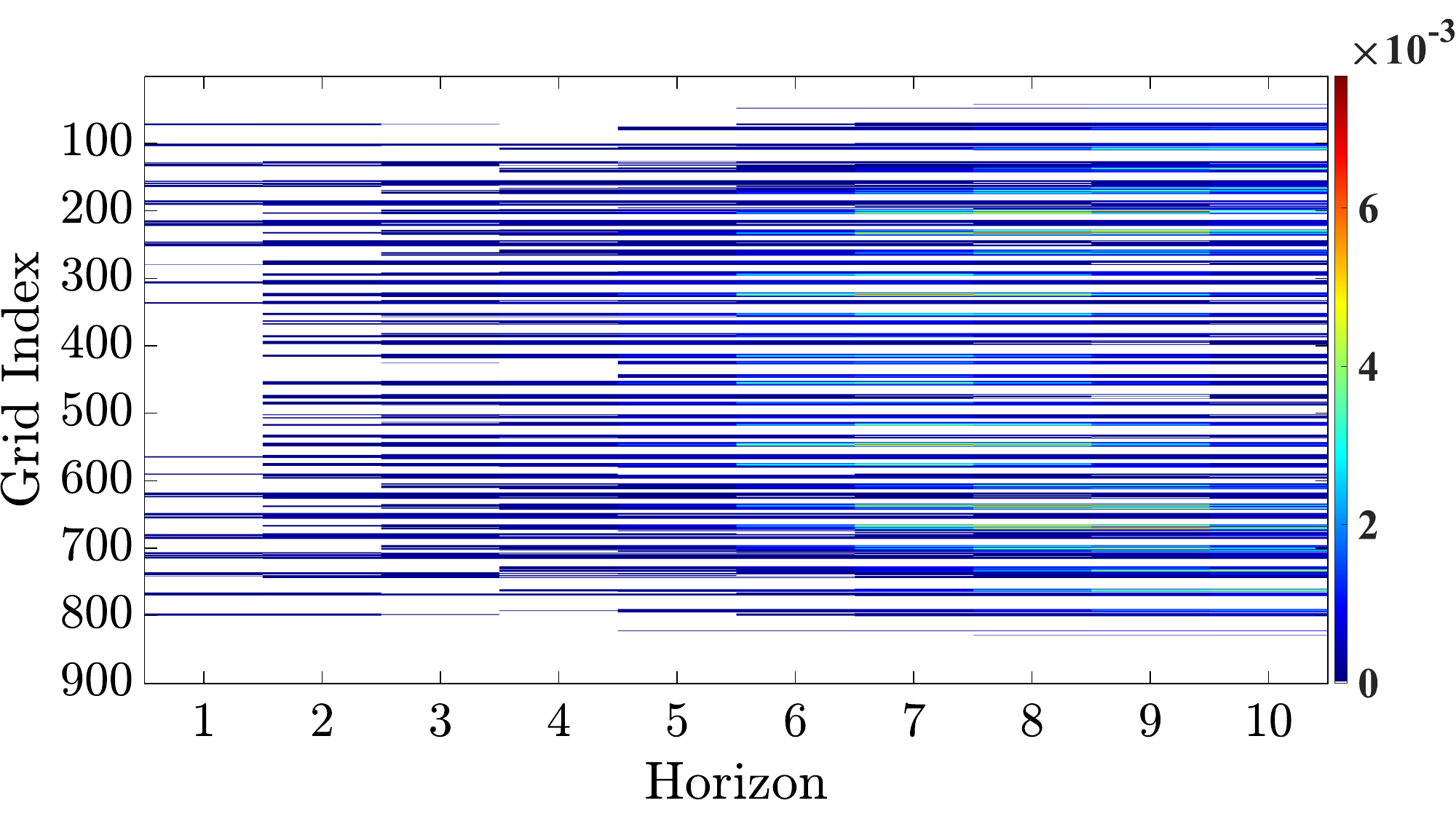}
	    \captionsetup{font=normalsize}
	    \caption{Distributed error heatmap showing the absolute prediction error of the distributed Koopman operator \( K_\mathrm{d} \) against the true system snapshots over the prediction horizon.}
	    \label{fig:error_distr_vs_true}
	\end{figure}

	 Fig.~\ref{fig:prediction_discrepancy} shows the evolution of the prediction discrepancy between distributed and centralized Koopman estimates, measured by the Frobenius norm:
	\begin{align} \label{mathcal__O}
	    \mathcal{O}(t) := \left\| (K^* - K_\mathrm{d}{(t)} )X   \right\|_F.
	\end{align}
	Here, \( K^* \) denotes the centralized Koopman operator derived from \eqref{eq:frobenius_minimization} and \( K_\mathrm{d}(t) \) the distributed estimate at iteration \( t \). The metric quantifies the difference in predicted lifted dynamics. As shown, \(\mathcal{O}(t) \) decays exponentially, confirming convergence to the centralized solution. The convergence rate depends on the step size, graph Laplacian spectrum, and matrix \( \bm{X} \), with \( \rho_{\max} \), validating the effectiveness of the distributed learning algorithm.
	Fig.~\ref{fig:K_diff} shows the entrywise absolute difference \( K_\Delta := |K_\mathrm{d} - K^*| \), with $[K_\Delta]_{ij} := |[K_\mathrm{d}]_{ij} - [K^*]_{ij}|.$
	The heatmap reveals that most entries have small errors, indicating strong alignment between distributed and centralized models. Block patterns reflect the partitioning across \( p=3 \) agents, and higher deviations appear sparsely, likely due to local variations or communication limits.
	Fig.~\ref{fig:eigenvalues} compares eigenvalues of \( K^* \) (blue) and \( K_\mathrm{d} \) (red) in the complex plane. All lie within the unit circle, confirming stability. \( K_\mathrm{d} \)'s spectrum is slightly more spread but closely matches dominant modes of \( K^* \), indicating successful preservation of key system dynamics via distributed learning.
	Fig.~\ref{fig:error_distr_vs_true} depicts the prediction error of \( K_\mathrm{d} \) over the whole spatial grid and prediction steps. Errors grow over time due to recursive forecasting, with some localized spikes likely from decentralized update limitations. This spatial-temporal map complements global metrics by revealing when and where prediction errors concentrate.

    \begin{remak}\label{rem:suboptimality}
	The results in Fig.~\ref{fig:prediction_discrepancy}--\ref{fig:error_distr_vs_true} demonstrate the progressive alignment of distributed Koopman estimates with the centralized reference solution over successive iterations. Initially, the estimate is suboptimal due to limited local information, but the consensus-based updates iteratively reduce this discrepancy, producing an exponential decay in the prediction error until near-equivalence with the centralized Koopman operator is achieved. The convergence rate is governed by the step size $\alpha$, the communication topology (via the Laplacian spectrum), and the conditioning of the lifted data matrices, encapsulated in $\rho_{\max}$.
	In online operation, the algorithm readily adapts to streaming or windowed data. When new observations become available, each agent updates its local matrices and resumes from the previous estimate rather than restarting from scratch \cite{zhang2019online}. This warm-start feature accelerates adaptation while preserving previously learned structure, ensuring that the suboptimality gap remains bounded and vanishes with continued iterations. Consequently, even intermediate solutions are sufficiently predictive for real-time deployment, while full consensus yields operators virtually identical to their centralized counterparts. This trade-off between adaptivity, convergence, and computational scalability makes the proposed approach suitable for continuous learning in dynamic, data-driven environments.
	\end{remak}


	In the simulations, the lifting is implemented via direct vectorization of the spatial grid, treating each pixel intensity as an observable. This non-parametric choice preserves all spatial information by aligning the lifted space with the perceptual domain, making it well suited for dynamic environments with unknown structure.
To assess the impact of lifting richness, we consider the following two performance metrics:
\[
\mathcal{Q}_1 := \frac{\|Y-K_\mathrm{d}(t_{\max})X\|_F}{\|Y\|_F},
~
\mathcal{Q}_2 := \frac{\|K^*-K_\mathrm{d}(t_{\max})\|_F}{\|K^*\|_F},
\]
in addition to the transient prediction discrepancy $\mathcal{O}(t)$ defined in~\eqref{mathcal__O}.  
For direct vectorization of local raw data segments, $\mathcal{O}(t)$ reaches $10^{-3}$ at approximately $163$ iterations, with $\mathcal{Q}_1$ on the order of $10^{-5}$ and $\mathcal{Q}_2\simeq0.43$.  
Augmenting the local lifting with first- and second-order powers yields comparable accuracy, with $\mathcal{O}(t)$ reaching $10^{-3}$ at around $170$ iterations, $\mathcal{Q}_1$ remaining on the order of $10^{-5}$, and $\mathcal{Q}_2\simeq0.50$.  
Including up to third-order powers further increases the lifted dimension but does not improve performance: $\mathcal{O}(t)$ reaches $10^{-3}$ after about $174$ iterations, while $\mathcal{Q}_1$ remains at the same order and $\mathcal{Q}_2$ increases to approximately $0.55$.  
These results indicate that, in this setting, enriching the lifting does not necessarily improve either prediction accuracy or convergence speed and may even degrade agreement with the centralized solution, while incurring higher per-iteration computational cost. This highlights that richer liftings are not universally beneficial and that lightweight, well-aligned representations can offer a more effective accuracy--efficiency trade-off in distributed Koopman learning.

	\begin{remak}\label{rem:alphaaaaaaa}
	The convergence behavior of the distributed Koopman learning algorithm is closely tied to the parameters $\rho_{\max}$ and $\alpha$. The value of $\rho_{\max}$ depends on spectral properties of both the data matrices (in particular, $\bm X$) and the communication graph Laplacian $L$, as seen in the expression of~\eqref{def:rho_max}. Intuitively, richer and more informative data snapshots, as well as well-connected network topologies, reduce $\rho_{\max}$ and improve convergence. The step-size $\alpha$ acts as a design parameter: larger values can accelerate convergence initially but risk instability if chosen beyond the admissible range dictated by $\rho_{\max}$, while smaller values ensure stability but slow down the learning process. In practice, $\alpha$ must be tuned to balance convergence speed and numerical robustness. These parameters thus directly influence both the transient prediction discrepancy (cf. Fig.~\ref{fig:prediction_discrepancy}) and the quality of the final Koopman approximation, especially in online settings where rapid adaptation to new data is critical.
	\end{remak}

\begin{remak}
In practical sensing scenarios, the fields of view of different UAV agents may partially overlap, leading to repeated measurements of the same physical grid cells across agents. Such overlap does not invalidate the distributed Koopman learning algorithm: convergence of Algorithm~\ref{algo} guarantees recovery of the optimal Koopman operator associated with the chosen lifted coordinate system, even if this representation contains redundant (duplicated) states. However, convergence alone does not automatically remove redundancy; if overlapping grid cells are included multiple times, the learned operator corresponds to a redundant but dynamically consistent coordinate representation.
In practice, redundancy can be handled in several standard and fully valid ways. A common approach is to impose a unique global indexing of spatial cells prior to learning, assigning each physical state to a single agent and thereby avoiding duplication altogether. Alternatively, overlapping coordinates may be retained during learning and subsequently fused through a linear aggregation or averaging map, yielding a reduced Koopman model defined on a unique global state space. Such post-processing steps preserve the learned dynamics while eliminating redundancy when a minimal representation is desired.
In the presented simulations, a non-overlapping spatial partition is adopted for clarity. Nevertheless, the proposed distributed framework naturally accommodates overlapping fields of view through the above mechanisms, making it directly applicable to realistic multi-UAV sensing configurations.
\end{remak}


    \subsection{Comparison with other distributed Koopman methods}
The proposed distributed Koopman learning algorithm is benchmarked against representative approaches from the literature under identical experimental conditions. For a fair comparison, the same dataset, lifting function, and data partitions were used for the methods in~\cite{liu2020towards} and~\cite{nandanoori2021data}. All algorithms operated on identical splits of the lifted data matrices $X$ and $Y$ and were evaluated using normalized one-step and multi-step prediction errors.

The normalized prediction metrics are defined as
\[
e_1 = \frac{\|Y - K_\mathrm{d}X\|_F}{\|Y - K^*X\|_F}, \qquad
e_h = \frac{\|Y^{(h)} - K_\mathrm{d}^h X\|_F}{\|Y^{(h)} - (K^*)^h X\|_F},
\]
where $Y^{(h)}$ denotes the $h$-step-ahead lifted data. Numerical results show that all distributed methods closely approximate the centralized Koopman operator, with comparable one-step errors ($e_1\approx1.01{-}1.02$). For longer horizons ($h=14$), corresponding to the MPC planning window, the proposed method achieves a $11\%-13\%$ reduction in multi-step prediction error relative to~\cite{liu2020towards,nandanoori2021data}. This improved long-horizon accuracy enhances the reliability of Koopman-based predictions for MPC, yielding smoother and safer navigation in dynamic environments.

    \subsection{MPC with Koopman-based dynamic obstacle avoidance}
After validating the distributed Koopman model, its predictions are embedded into an MPC framework as time-varying polytopic constraints, enabling safe and goal-directed robot navigation over the prediction horizon.

            	\begin{figure}[!t]
	    \centering
    \includegraphics[trim=3.0cm 0.0cm 3.0cm 0.0cm, clip, width=0.45\textwidth]{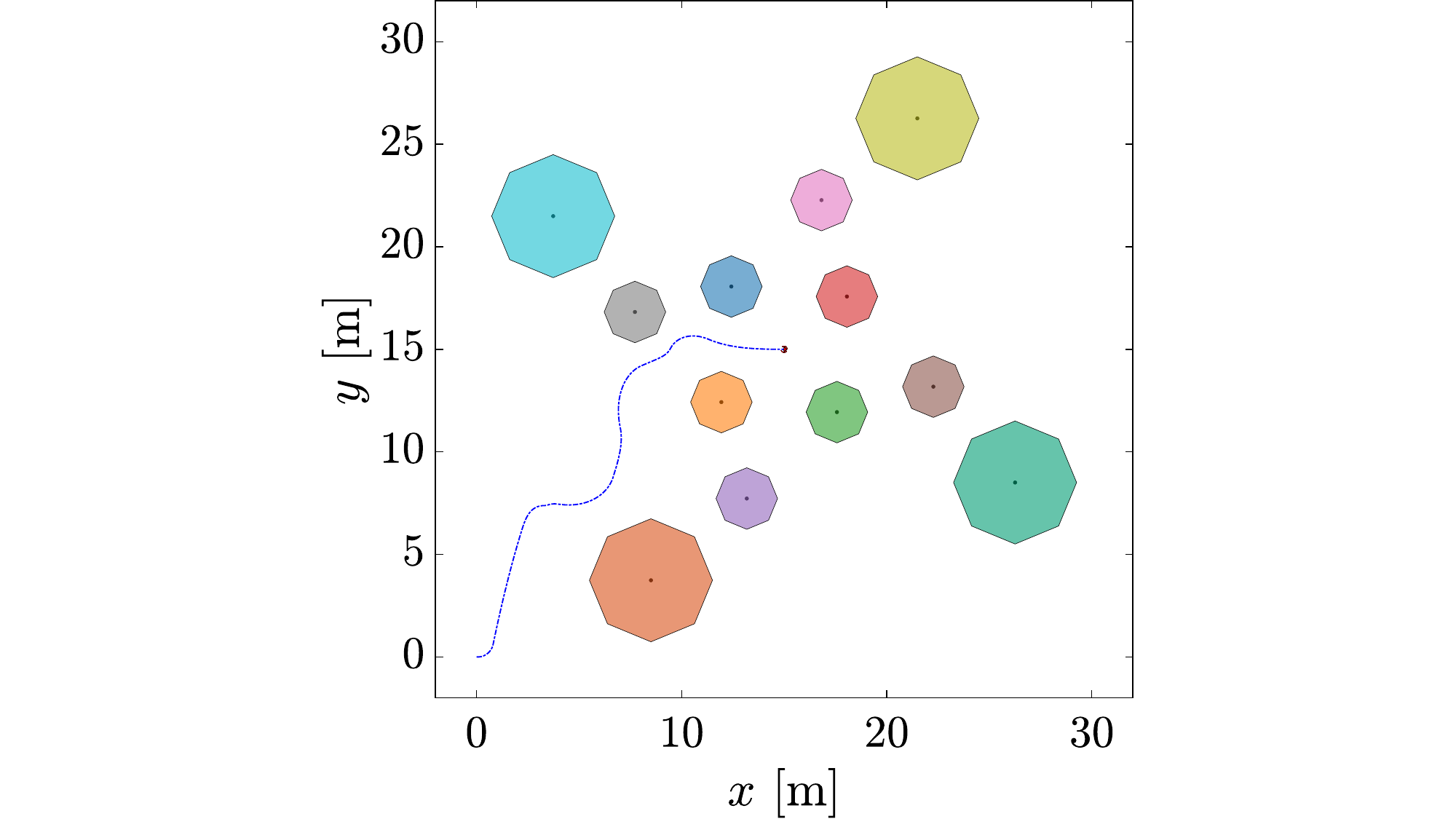}
	    \captionsetup{font=normalsize}
	    \caption{The final robot trajectory with Koopman-predicted dynamic obstacle avoidance.}
	    \label{fig:robot_traj_prediction}
	\end{figure}
       	\begin{figure}[!t]
	\centering
	\subfigure[$\mathrm{Time = 15~[\mathrm{s}]}$.]{
	\includegraphics[trim=3.0cm 0.0cm 3.0cm 0.0cm, clip, width=0.45\textwidth]{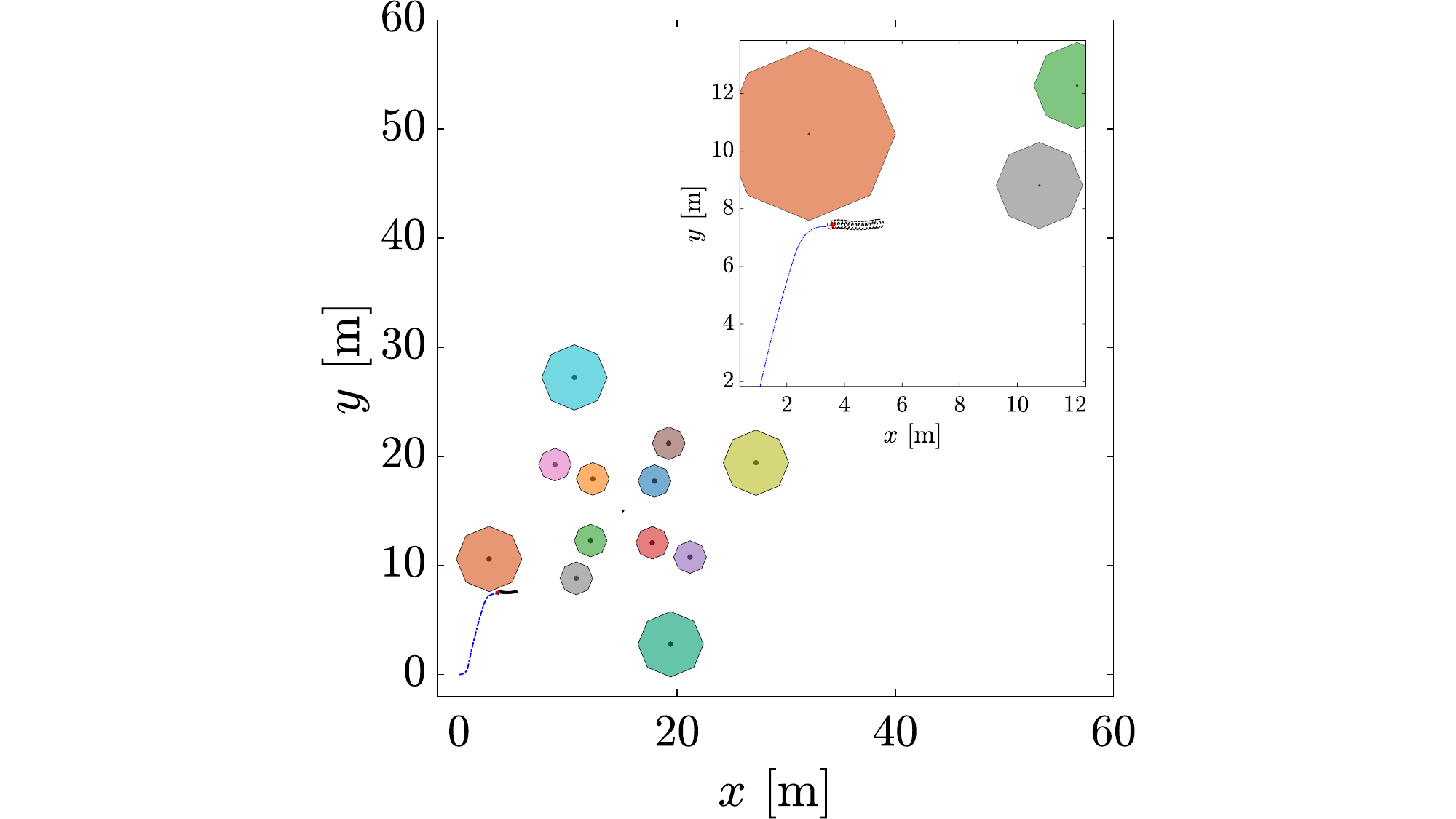}\label{fig:sim11xxx}
	}
	\subfigure[$\mathrm{Time = 20~[\mathrm{s}]}$.]{
	\includegraphics[trim=3.0cm 0.0cm 3.0cm 0.0cm, clip, width=0.45\textwidth]{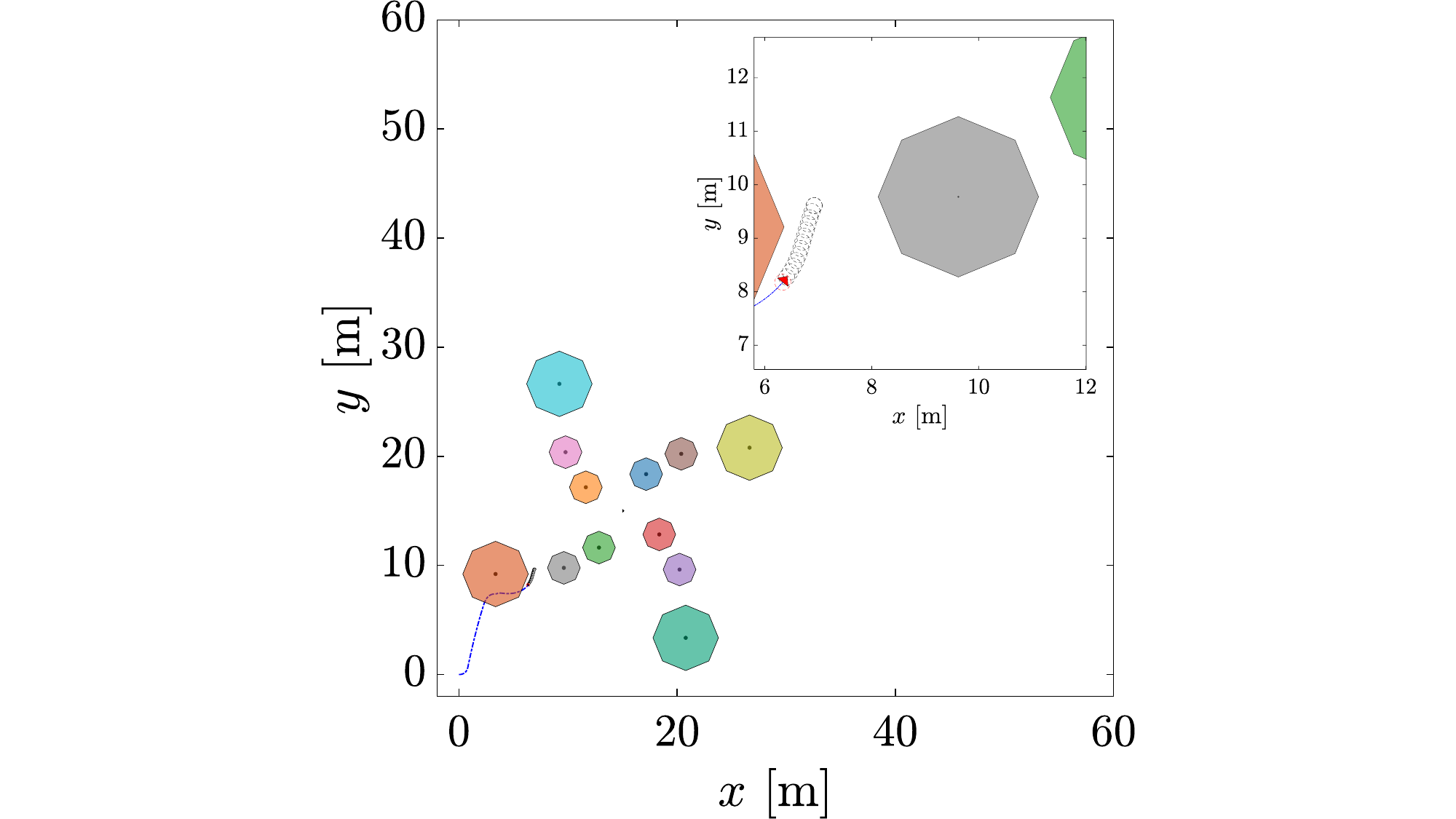}\label{fig:sim22xxx}
	}
	\captionsetup{font=normalsize}
	\caption{Snapshots of robot and obstacle positions at different time instances.}
	\label{fig:snapsxxx}
	\end{figure}
    
	The cost function in~\eqref{eq:mpc_cost} employs a prediction horizon $H=14$, initial position $\mathrm{z}(0) = [0,\, 0]^\top$, and target $\mathrm{z}^* = [15,\, 15]^\top$. 
	Dynamic obstacle avoidance is incorporated by approximating each Koopman-predicted elliptical region with a convex polytope defined by ${n_{\varsigma}} = 8$ supporting hyperplanes as an effective trade-off between accuracy and computational cost. The linear inequalities of the form~\eqref{OBSTACLEFINAL} are imposed at each horizon step, enforcing only the most active constraint per obstacle to ensure real-time feasibility while maintaining conservatism.
	
\begin{figure}[!t]
	    \centering
	       \includegraphics[trim=0.0cm 0.0cm 0.0cm 0.0cm, clip, width=0.45\textwidth]{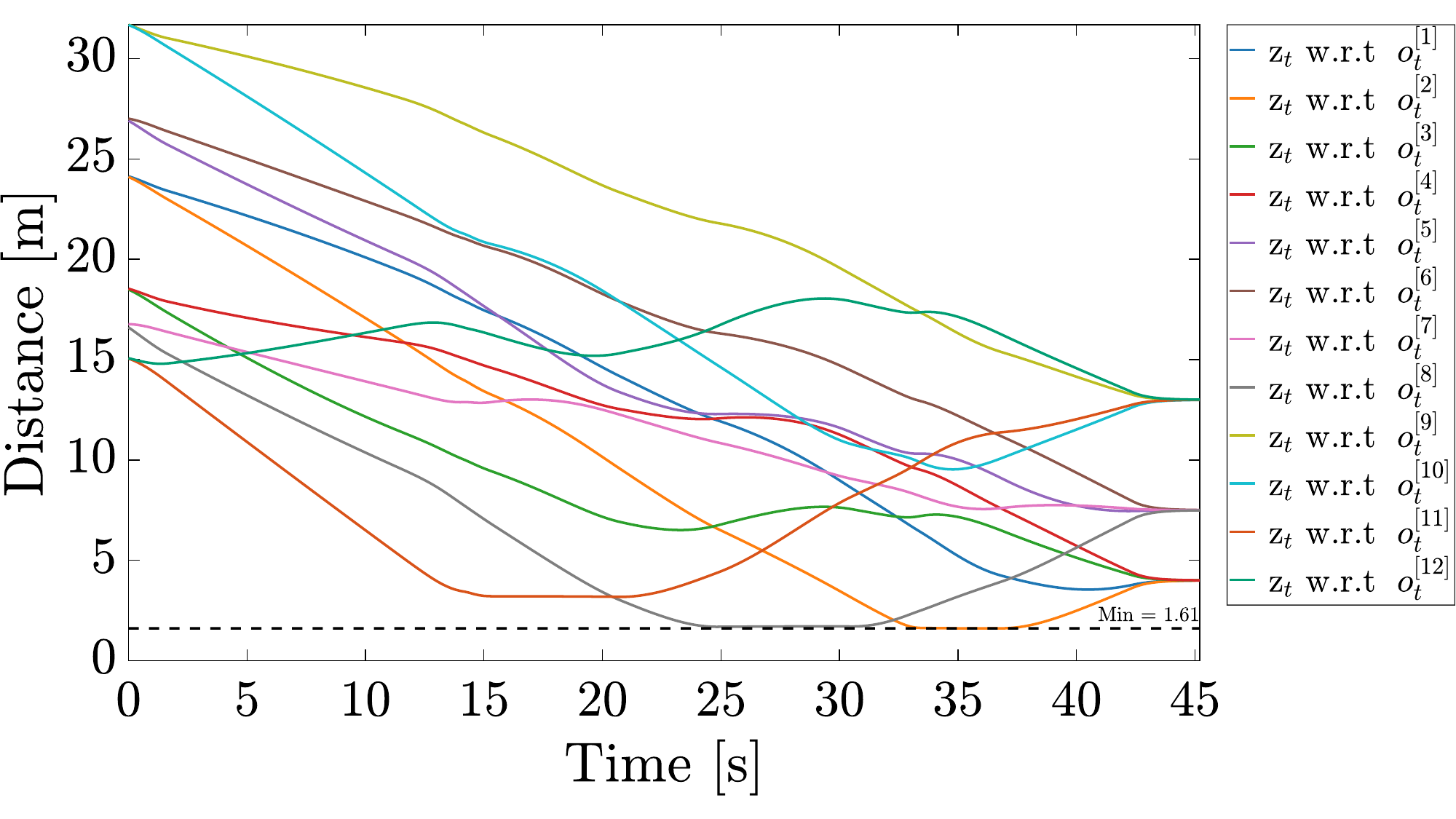}
	     \captionsetup{font=normalsize}
	    \caption{Robot-to-obstacle distance over time.}
	    \label{fig:robot_obstacle_distances}
	\end{figure}

\begin{figure}[!b]
	    \centering
	       \includegraphics[trim=0.0cm 0.0cm 0.0cm 0.0cm, clip, width=0.45\textwidth]{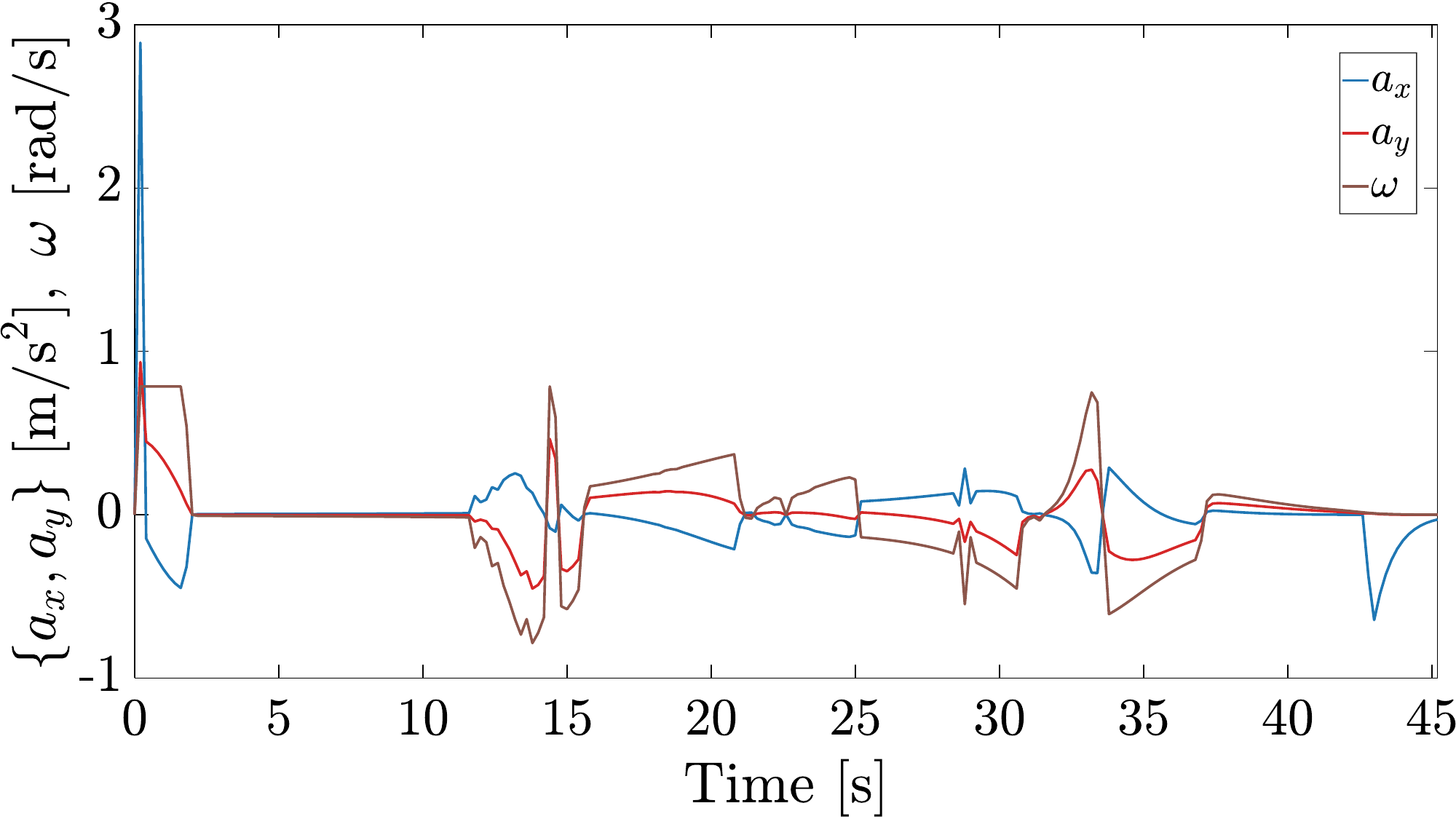}
	     \captionsetup{font=normalsize}
	    \caption{Evolution of control inputs.}
	    \label{fig:control_inputs_profile}
	\end{figure}

	Fig.~\ref{fig:robot_traj_prediction} shows the closed-loop trajectory in an environment with $L_{o}=12$ moving obstacle clusters where we observe that the robot reaches its target smoothly, maintaining safe separation by continuously adapting to predicted obstacle motion. Fig.~\ref{fig:snapsxxx} illustrates representative navigation snapshots, showing the robot’s orientation, predicted trajectory, and evolving obstacle regions, and confirming anticipatory maneuvering enabled by Koopman-based forecasts. Fig.~\ref{fig:robot_obstacle_distances} reports the Euclidean distance between the robot and obstacle centers, which consistently remains above the safety threshold, verifying robust collision avoidance. Fig.~\ref{fig:control_inputs_profile} shows the corresponding control inputs, where translational accelerations $(a_x,a_y)$ and angular velocity $\omega$ yield dynamically feasible and coordinated motion. Together, these results demonstrate safe and predictive navigation achieved through integration of Koopman-based predictions into MPC. For implementation details, see~\cite{bueno2025koopman}.



   \begin{remak}\label{rem:scalability}
Scalability of the proposed framework arises from both distributed Koopman learning and its MPC integration. During learning, row-wise partitioning of the lifted data ensures that each agent processes only local blocks $X_i,Y_i$, reducing per-agent memory to ${O}(n_iN)$ and computation to ${O}(n_i^2N)$ per iteration, which remains constant under weak scaling as the number of agents grows. Communication overhead is limited to exchanging compact variables ($S_i$), making bandwidth requirements dependent on graph connectivity rather than data dimension. Convergence is governed by the contraction factor $\rho_{\max}$, leading to near-linear wall-clock speedup with additional agents.  
In the MPC stage, Koopman-based obstacle predictions are embedded as linear constraints using polygonal approximations with ${n_{\varsigma}}=8$ facets. Activating only the most restrictive hyperplane per obstacle and prediction step reduces constraint complexity from ${O}(L_o H n_{\varsigma})$ to ${O}(L_o H)$ while preserving conservative safety margins. The resulting sparse linear constraints maintain MPC tractability and enable efficient warm-starting, supporting real-time operation in high-dimensional dynamic environments.
\end{remak}

    	\begin{figure}[!t]
	    \centering
	      \includegraphics[trim=3.0cm 0.0cm 3.0cm 0.0cm, clip, width=0.45\textwidth]{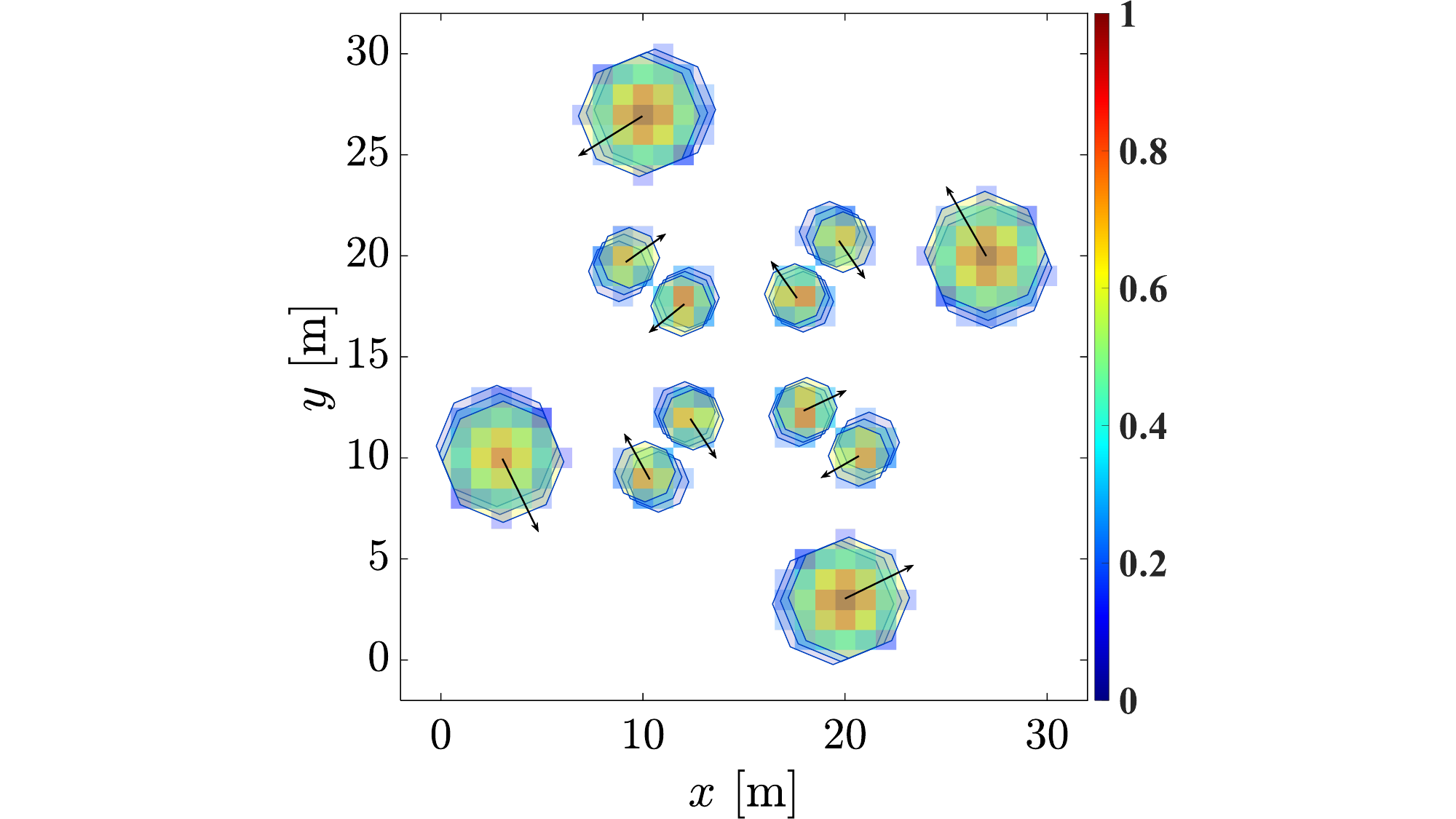}
	     \captionsetup{font=normalsize}
	    \caption{Predicted intensity–polytope overlays at representative time and its corresponding future horizons.}
	    \label{fig:intense11xxx}
	\end{figure}

	\subsection{Linking Koopman predictions to MPC constraints}

To clarify how perception informs planning, we visualize the predicted spatial intensities \(\rho_{t+h}\) generated by the distributed Koopman operator together with the geometric obstacle sets used by MPC at representative times corresponding to Fig.~\ref{fig:sim11xxx}. From the most recent lifted snapshot at time \(t\), predictions are obtained via \(X_{t+h}\approx K^{h}X_t\), \(h=1,\dots,H\), and mapped back to the grid to recover \(\rho_{t+h}\in[0,1]\). A threshold \(c_{\rho}\) selects high-likelihood regions \(\Psi_t=\{\xi\in\mathbb{R}^2:\rho_t(\xi)>c_{\rho}\}\), to which a Gaussian mixture model is fitted, yielding component means \(\{\mu_{\ell,t+h}\}\) and covariances \(\{\Sigma_{\ell,t+h}\}\). Confidence ellipses derived from \((\mu_{\ell,t+h},\Sigma_{\ell,t+h})\) are then polygonally approximated and inflated by a safety margin \(\varepsilon\), producing time-varying polytopes that conservatively bound the predicted obstacle mass.
Fig.~\ref{fig:intense11xxx} overlays \(\rho_{t+h}\) at \(t=15\) for \(h=1,7,14\), highlighting intensity values within the union of all nominal polytopes, which are passed to MPC as the linear constraints in \eqref{OBSTACLEFINAL}. This visualization illustrates three key links: \emph{(i)} the Koopman operator predicts future obstacle distributions on the original grid, \emph{(ii)} the Gaussian mixture model provides a compact parametric surrogate of \(\rho_t\), and \emph{(iii)} polygonal circumscribers convert these probabilistic predictions into conservative linear inequalities enforceable along the MPC horizon. As time evolves, the intensity ridges advect and deform, the mixture components re-center and re-orient, and the resulting polytopes, constructed with \(n_{\varsigma}=8\) facets and inflated boundaries, define the actual constraints enforced in \eqref{OsbtacleMPC}. This pipeline explains the anticipatory behavior of the robot: future occupancy is forecast through \(\rho_{t+h}\), summarized by polytopes, and incorporated into MPC before high-density regions intersect the planned path.

	\section{Conclusion}
\label{Sec. 5}

This paper presented a unified data-driven framework that integrates distributed Koopman operator learning with MPC for predictive navigation and dynamic obstacle avoidance. High-dimensional sensory data are exploited to collaboratively learn and forecast obstacle motion without centralized data fusion. The resulting Koopman-based predictions, represented via Gaussian mixtures and polytope approximations, are incorporated as linear constraints in the MPC formulation, enabling safe and collision-free navigation under uncertainty.
The proposed approach is inherently scalable and well suited for autonomous navigation in complex, time-varying environments. Distributed learning ensures computational efficiency, limited communication overhead, and real-time adaptability. Convergence of the distributed Koopman estimator is theoretically guaranteed, and simulation results validate reliable, safe, and scalable motion planning. Overall, this work advances cooperative perception and distributed control for networked autonomous systems by tightly integrating learning-based modeling with predictive control.

	\section{Acknowledgment}
	The authors are thankful to Marcello Farina for several enlightening discussions on the proposed approach.
	
	\balance
	\bibliographystyle{IEEEtran}       	        
	\bibliography {biblio}
	
	\end{document}